\documentclass[sigconf]{acmart}
\PassOptionsToPackage{bookmarks=false}{hyperref}
\setcopyright{acmcopyright}
\acmConference[ICSE '18]{ICSE '18: 40th International Conference on Software Engineering }{May 27-June 3, 2018}{Gothenburg, Sweden}

\usepackage{paralist}
\usepackage{etoolbox}
\usepackage{microtype}
\usepackage[bookmarks=false]{hyperref}

\usepackage{fancybox}
\usepackage{graphicx}
\graphicspath{{figures/}}
\DeclareGraphicsExtensions{.pdf,.jpeg,.png}
\usepackage{amsmath}
\usepackage{subfigure}
\usepackage{caption}
\usepackage{fancyvrb}
\usepackage{color}
\usepackage{algorithmic}
\PassOptionsToPackage{hyphens}{url}\usepackage{hyperref}
\usepackage[linesnumbered, ruled]{algorithm2e}
\usepackage{array}
\usepackage{mdwmath}
\usepackage{mdwtab}
\usepackage{booktabs}
\usepackage{listings}
\usepackage{eqparbox}
\usepackage{stfloats}
\usepackage{url}
\usepackage{comment}
\usepackage{framed}
\usepackage{ifthen}
\usepackage{balance}
\usepackage{enumitem}
\usepackage{ragged2e}
\usepackage{stackengine}
\usepackage{listofitems}
\usepackage[flushleft]{threeparttable}
\usepackage{makecell}
\usepackage[turnon]{notes}
\usepackage{multirow}
\usepackage{xspace}
\usepackage{listings}
\usepackage{bm}
\usepackage[export]{adjustbox}

\newcommand{\rom}[1]{\uppercase\expandafter{\romannumeral #1\relax}}
\newcommand{\pname}{DeepTest\xspace}

\newcommand{\etal}{\hbox{\emph{et al.}}\xspace}
\newcommand{\eg}{\hbox{\emph{e.g.,}}\xspace}
\newcommand{\ie}{\hbox{\emph{i.e.}}\xspace}

\newcommand{\wrt}{\hbox{\emph{w.r.t.}}\xspace}

\newcommand{\etc}{\hbox{\emph{etc.}}\xspace}

\def\wu{$^{\dag}$}

\definecolor{gray50}{gray}{.5}
\definecolor{gray40}{gray}{.6}
\definecolor{gray30}{gray}{.7}
\definecolor{gray20}{gray}{.8}
\definecolor{gray10}{gray}{.9}
\definecolor{gray05}{gray}{.95}

\newlength\Linewidth
\def\findlength{\setlength\Linewidth\linewidth
\addtolength\Linewidth{-4\fboxrule}
\addtolength\Linewidth{-3\fboxsep}
}
\newenvironment{examplebox}{\par\begingroup
   \setlength{\fboxsep}{5pt}\findlength
   \setbox0=\vbox\bgroup\noindent
   \hsize=0.95\linewidth
   \begin{minipage}{0.95\linewidth}\normalsize}
    {\end{minipage}\egroup
    \textcolor{gray20}{\fboxsep1.5pt\fbox
     {\fboxsep5pt\colorbox{gray05}{\normalcolor\box0}}}
    \endgroup\par\noindent
    \normalcolor\ignorespacesafterend}

\newcounter{RQCounter}
\newcounter{RQACounter}

\newcommand{\RQ}[2]{%
\refstepcounter{RQCounter} \label{#1}
 \begin{center}	
  \begin{examplebox}
   \textbf{RQ\arabic{RQCounter}.}~#2
  \end{examplebox}	 
 \end{center}
}

\newcommand{\RS}[2]{%
\begin{framed}%
\filbreak
\textbf{Result {\ref{#1}}:~}{\emph {#2}}%
\end{framed}
}

\definecolor{javared}{rgb}{0.6,0,0} 
\definecolor{javagreen}{rgb}{0.25,0.5,0.35} 
\definecolor{javapurple}{rgb}{0.5,0,0.35} 
\definecolor{javadocblue}{rgb}{0.25,0.35,0.75} 

\lstdefinestyle{customc}{
  belowcaptionskip=\baselineskip,
  breaklines=true,
  xleftmargin=\parindent,
  language=java,
  showstringspaces=false,
  basicstyle=\scriptsize\ttfamily,
  keywordstyle=\bfseries\color{javapurple},
  commentstyle=\itshape\blue,
}

\DeclareCaptionLabelFormat{andimage}{#1~#2  \&  \figurename~\thefigure}

\newcommand{\rambo}{Rambo~}
\newcommand{\rs}{Rambo-S1~}
\newcommand{\rt}{Rambo-S2~}
\newcommand{\ru}{Rambo-S3~}
\newcommand{\chauf}{Chauffeur~}
\newcommand{\epoch}{Epoch~}
\newcommand{\cnn}{Chauffeur-CNN~}
\newcommand{\lstm}{Chauffeur-LSTM~}

\newcommand{\positive}{left (+ve)~}
\newcommand{\negative}{right (-ve)~}

\newcommand{\iorg}{$I_{o}$}
\newcommand{\itrans}{$I_{t}$}

\newcommand{\isusp}{$I_{err}$}
\newcommand{\iorig}{$I_{org}$}

\newcommand{\Comment}[1]{}

\newcommand\Red[1]{\textcolor[rgb]{1.00,0.00,0.00}{\textbf{#1}}}

\newcommand\blue[1]{\textcolor[rgb]{0.00,0.00,1.00}{{#1}}}

\usepackage{epstopdf}
\lstdefinestyle{example} {
	frame=tb,
	basicstyle=\footnotesize\ttfamily,
	numbers=left,
	language=C
}

\begin{document}\sloppy

%

\copyrightyear{2018} 
\acmYear{2018} 
\setcopyright{acmcopyright}
\acmConference[ICSE '18]{ICSE '18: 40th International Conference on Software Engineering }{May 27-June 3, 2018}{Gothenburg, Sweden}
\acmBooktitle{ICSE '18: ICSE '18: 40th International Conference on Software Engineering , May 27-June 3, 2018, Gothenburg, Sweden}
\acmPrice{15.00}
\acmDOI{10.1145/3180155.3180220}
\acmISBN{978-1-4503-5638-1/18/05}

\title{DeepTest: Automated Testing of \\ Deep-Neural-Network-driven Autonomous Cars}


\author{Yuchi Tian}
\affiliation{%
\institution{University of Virginia}
}
\email{yuchi@virginia.edu}

\author{Kexin Pei}
\affiliation{%
\institution{Columbia University}
}
\email{kpei@cs.columbia.edu}

\author{Suman Jana}
\affiliation{%
\institution{Columbia University}
}
\email{suman@cs.columbia.edu}

\author{Baishakhi Ray}
\affiliation{%
\institution{University of Virginia}
}
\email{rayb@virginia.edu}


\makeatletter
\patchcmd{\@maketitle}
  {\addvspace{0.1\baselineskip}\egroup}
  {\addvspace{-2\baselineskip}\egroup}
  {}
  {}


\makeatother

\renewcommand{\shortauthors}{Tian et al.}

\begin{abstract}

Recent advances in Deep Neural Networks (DNNs) have led to the development of DNN-driven autonomous cars that, using sensors like camera, LiDAR, etc., can drive without any human intervention.  Most major manufacturers including Tesla, GM, Ford, BMW, and Waymo/Google are working on building and testing different types of autonomous vehicles. The lawmakers of several US states including California, Texas, and New York have passed new legislation to fast-track the process of testing and deployment of autonomous vehicles on their roads.

However, despite their spectacular progress, DNNs, just like traditional software, often demonstrate incorrect or unexpected corner-case behaviors that can lead to potentially fatal collisions. Several such real-world accidents involving autonomous cars have already happened including one which resulted in a fatality. Most existing testing techniques for DNN-driven vehicles are heavily dependent on the manual collection of test data under different driving conditions which become prohibitively expensive as the number of test conditions increases.  

In this paper, we design, implement, and evaluate \pname, a systematic testing tool for automatically detecting erroneous behaviors of DNN-driven vehicles that can potentially lead to fatal crashes. First, our tool is designed to automatically generated test cases leveraging real-world changes in driving conditions like rain, fog, lighting conditions, etc. \pname systematically explore different parts of the DNN logic by generating test inputs that maximize the numbers of activated neurons. \pname found thousands of erroneous behaviors under different realistic driving conditions (e.g., blurring, rain, fog, etc.) many of which lead to potentially fatal crashes in three top performing DNNs in the Udacity self-driving car challenge.

\end{abstract}



\maketitle





\section{Introduction}
\label{sec:intro}

Significant progress in Machine Learning (ML) techniques like Deep Neural Networks (DNNs) over the last decade has enabled the development of safety-critical ML systems like autonomous cars.
Several major car manufacturers including Tesla, GM, Ford, BMW, and Waymo/Google are building and actively testing these cars. Recent results show that autonomous cars have become very efficient in practice and already driven millions of miles without any human intervention~\cite{selfdrivingcar1, selfdrivingcar2}. Twenty US states including California, Texas, and New York have recently passed legislation to enable testing and deployment of autonomous vehicles~\cite{selfdrivingcar4}.

However, despite the tremendous progress, just like traditional software, DNN-based software, including the ones used for autonomous driving, often demonstrate incorrect/unexpected corner-case behaviors that can lead to dangerous consequences like a fatal collision. Several such real-world cases have already been reported (see Table~\ref{tab:accident}). As Table~\ref{tab:accident} clearly shows, such crashes often happen under rare previously unseen corner cases. For example, the fatal Tesla crash resulted from a failure to detect a white truck against the bright sky. The existing mechanisms for detecting such erroneous behaviors depend heavily on manual collection of labeled test data or ad hoc, unguided simulation~\cite{waymoreport, waymo_simulation} and therefore miss numerous corner cases. Since these cars adapt behavior based on their environment as measured by different sensors (e.g., camera, Infrared obstacle detector, etc.),  the space of possible inputs is extremely large. Thus,  unguided simulations are highly unlikely to find many erroneous behaviors.


\begin{table*}[!htbp]
  \centering
  \setlength{\tabcolsep}{6pt}
    \scriptsize
    \renewcommand{\arraystretch}{.9}
  \caption{\textbf{\small Examples of real-world accidents involving autonomous cars}}
    \begin{tabular}{lllp{2cm}p{7cm}}
         & \textbf{Reported Date} & \textbf{Cause} & \textbf{Outcome} & \textbf{Comments} \\ \toprule
    Hyundai Competition~\cite{hyundai.crash} & December, 2014 & Rain fall & Crashed while testing & "The sensors failed to pick up street signs, lane markings, and even pedestrians due to the angle of the car shifting in rain and the direction of the sun"~\cite{hyundai.crash}\\ \midrule
   
    Tesla autopilot mode~\cite{cnn.crash} & July, 2016 & Image contrast & Killed the driver & "The camera failed to recognize the white truck against a bright sky"~\cite{nyt.crash} \\ \midrule

    Google self-driving car~\cite{google.crash} & February, 2016 & Failed to estimate speed & Hit a bus while shifting lane & "The car assumed that the bus would yield when it attempted to merge back into traffic"~\cite{google.crash} \\ \bottomrule
    \end{tabular}%

  \label{tab:accident}%
\vspace{-0.4cm}
\end{table*}%

At a conceptual level, these erroneous corner-case behaviors in DNN-based software are analogous to logic bugs in traditional software. Similar to the bug detection and patching cycle in traditional software development, the erroneous behaviors of DNNs, once detected, can be fixed by adding the error-inducing inputs to the training data set and also by possibly changing the model structure/parameters. However, this is a challenging problem, as noted by large software companies like Google and Tesla that have already deployed machine learning techniques in several production-scale systems including self-driving car, speech recognition, image search, etc.~\cite{sculley2014machine,software2}.


Our experience with traditional software has shown that it is hard to build robust safety-critical systems only using manual test cases. Moreover, the internals of traditional software and new DNN-based software are fundamentally different. For example, unlike traditional software where the program logic is manually written by the software developers, DNN-based software automatically learns its logic from a large amount of data with minimal human guidance. In addition, the logic of a traditional program is expressed in terms of control flow statements while DNNs use weights for edges between different neurons and nonlinear activation functions for similar purposes. These differences make automated testing of DNN-based software challenging by presenting several interesting and novel research problems. 

First, traditional software testing techniques for systematically exploring different parts of the program logic by maximizing branch/code coverage is not very useful for DNN-based software as the logic is not encoded using control flow~\cite{pei2017deepxplore}. Next, DNNs are fundamentally different from the models (e.g., finite state machines) used for modeling and testing traditional programs. Unlike the traditional models, finding inputs that will result in high model coverage in a DNN is significantly more challenging due to the non-linearity of the functions modeled by DNNs. Moreover, the Satisfiability Modulo Theory (SMT) solvers that have been quite successful at generating high-coverage test inputs for traditional software are known to have trouble with formulas involving floating-point arithmetic and highly nonlinear constraints, which are commonly used in DNNs. In fact, several research projects have already attempted to build custom tools for formally verifying safety properties of DNNs. Unfortunately, none of them scale well to real-world-sized DNNs~\cite{katz2017reluplex, huang2017safety, pulina2010abstraction}. Finally, manually creating specifications for complex DNN systems like autonomous cars is infeasible as the logic is too complex to manually encode as it involves mimicking the logic of a human driver.

In this paper, we address these issues and design a systematic testing methodology for automatically detecting erroneous behaviors of DNN-based software of self-driving cars. First, we leverage the notion of neuron coverage (i.e., the number of neurons activated by a set of test inputs) to systematically explore different parts of the DNN logic. We empirically demonstrate that changes in neuron coverage are statistically correlated with changes in the actions of self-driving cars (e.g., steering angle). Therefore, neuron coverage can be used as a guidance mechanism for systemically exploring different types of car behaviors and identify erroneous behaviors. Next, we demonstrate that different image transformations that mimic real-world differences in driving conditions like changing contrast/brightness, rotation of the camera result in activation of different sets of neurons in the self-driving car DNNs. We show that by combining these image transformations, the neuron coverage can be increased by $100$\% on average compared to the coverage achieved by manual test inputs. Finally, we use transformation-specific metamorphic relations between multiple executions of the tested DNN (e.g., a car should behave similarly under different lighting conditions) to automatically detect erroneous corner case behaviors. We found thousands of erroneous behaviors across the three top performing DNNs in the Udacity self-driving car challenge~\cite{challenge2}.

\begin{figure}[!htb]
\centering
\subfigure[original]{\label{fig:baseline3}\includegraphics[width=0.4\columnwidth]{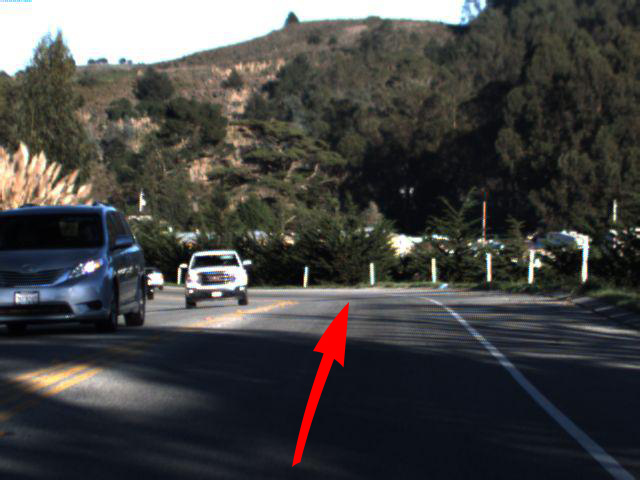}}
\qquad
\subfigure[with added rain]{\label{fig:rain3}\includegraphics[width=0.4\columnwidth]{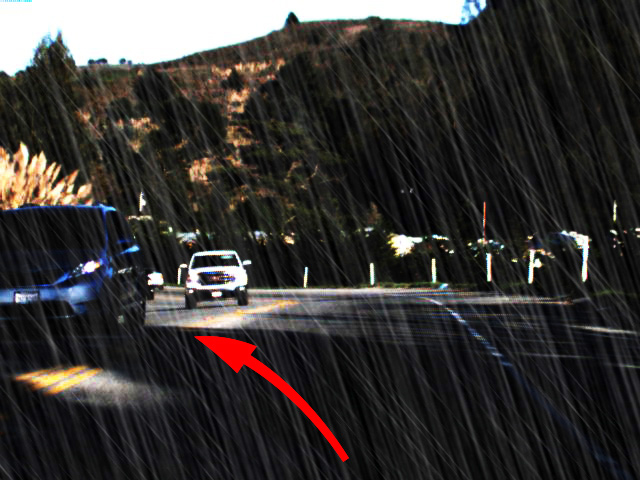}}
\vspace{-0.2in}
\caption{\textbf{\small A sample dangerous erroneous behavior found by \pname in the \textit{Chauffeur} DNN.}}
\label{fig:fix1}
\vspace{-0.4in}
\end{figure}

\noindent
The key contributions of this paper are:

\begin{itemize}[leftmargin=*]

\item
We present a systematic technique to automatically synthesize test cases that maximizes neuron coverage in safety-critical DNN-based systems like autonomous cars. We empirically demonstrate that changes in neuron coverage correlate with changes in an autonomous car's behavior. 
\item 
We demonstrate that different realistic image transformations like changes in contrast, presence of fog, etc. can be used to generate synthetic tests that increase neuron coverage. We leverage transformation-specific metamorphic relations to automatically detect erroneous behaviors.  Our experiments also show that the synthetic images can be used for retraining and making DNNs more robust to different corner cases.  
\item
We implement the proposed techniques in \pname, to the best of our knowledge, the first systematic and automated testing tool for DNN-driven autonomous vehicles. We use \pname to systematically test three top performing  DNN models from the Udacity driving challenge. \pname found thousands of erroneous behaviors in these systems many of which can lead to potentially fatal collisions as shown in Figure~\ref{fig:fix1}.  
\item 
We have made the erroneous behaviors detected by \pname available at \url{https://deeplearningtest.github.io/deepTest/}. We also plan to release the generated test images and the source of \pname for public use. 
\end{itemize}


\ \\

\section{Background}
\label{sec:model}

\subsection{Deep Learning for Autonomous Driving}
\label{subsec:dl_background}
\begin{figure}[!t]
\centering
\includegraphics[width=.95\columnwidth]{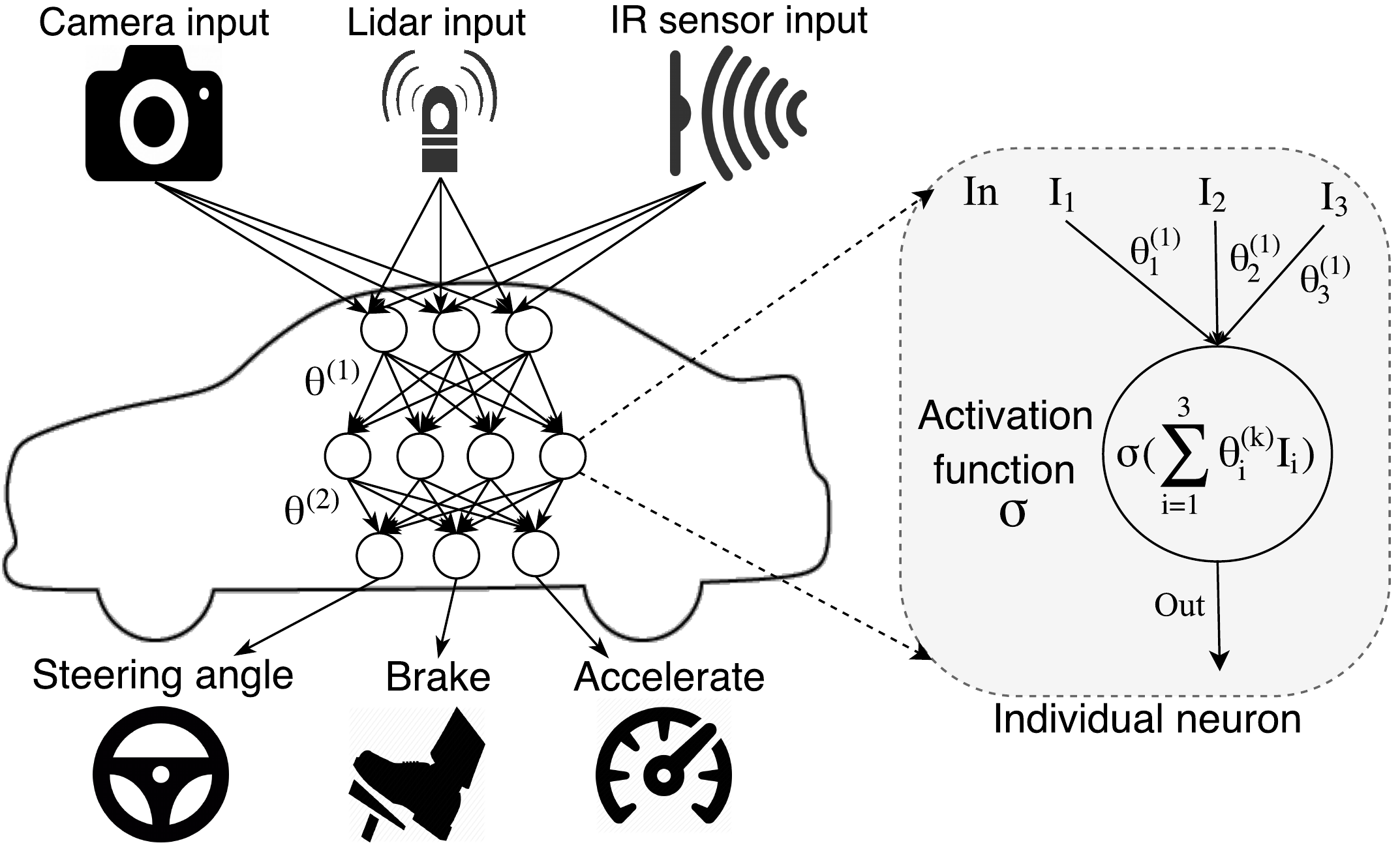}
\caption{\small A simple autonomous car DNN that takes inputs from camera, light detection and ranging sensor (LiDAR), and IR (infrared) sensor, and outputs steering angle, braking decision, and acceleration decision. The DNN shown here essentially models the function $\sigma(\bm{\theta^{(2)}}\cdot \sigma(\bm{\theta^{(1)}}\cdot \bm{x}))$ where $\bm{\theta}$s represent the weights of the edges and $\sigma$ is the activation function. The details of the computations performed inside a single neuron are shown on the right.}
\label{fig:driving_dnn}
\vspace{-0.4cm}
\end{figure}

The key component of an autonomous vehicle is the perception module controlled by the underlying Deep Neural Network (DNN)~\cite{apollo,autopilot}.
The DNN takes input from different sensors like camera, light detection and ranging sensor (LiDAR), and IR (infrared) sensor that measure the environment and outputs the steering angle, braking, etc. necessary to maneuver the car safely under current conditions as shown in Figure~\ref{fig:driving_dnn}. 
In this paper, we focus on the camera input and the steering angle output.


A typical feed-forward DNN is composed of multiple processing \textit{layers} stacked together to extract different representations of the input~\cite{bengio2007greedy}. Each layer of the DNN increasingly abstracts the input, e.g., from raw pixels to semantic concepts. For example, the first few layers of an autonomous car DNN extract low-level features such as edges and directions, while the deeper layers identify objects like stop signs and other cars, and the final layer outputs the steering decision (e.g., turning left or right). 

Each layer of a DNN consists of a sequence of individual computing units called \textit{neurons}. The neurons in different layers are connected with each other through edges. Each edge has a corresponding weight ($\bm{\theta}$s in Figure~\ref{fig:driving_dnn}). Each neuron applies a nonlinear \textit{activation function} on its inputs and sends the output to the subsequent neurons as shown in Figure~\ref{fig:driving_dnn}. Popular activation functions include ReLU (Rectified Linear Unit)~\cite{nair2010rectified}, sigmoid~\cite{Mitchell:1997:ML:541177}, etc. The edge weights of a DNN is inferred during the training process of the DNN based on labeled training data. Most existing DNNs are trained with gradient descent using backpropagation~\cite{rumelhart1988learning}. Once trained, a DNN can be used for prediction without any further changes to the weights. For example, an autonomous car DNN can predict the steering angle based on input images.


Figure~\ref{fig:driving_dnn} illustrates a  basic DNN in the perception module of a self-driving car. Essentially, the DNN is a sequence of linear transformations (\eg dot product between the weight parameters $\bm{\theta}$ of each edge and the output value of the source neuron of that edge) and nonlinear activations (e.g., ReLU in each neuron). Recent results have demonstrated that a well-trained DNN $f$ can predict the steering angle with an accuracy close to that of a human driver~\cite{bojarski2016end}.


\subsection{Different DNN Architectures}
Most DNNs used in autonomous vehicles can be categorized into two types: (1) Feed-forward Convolutional Neural Network (CNN), and (2) Recurrent neural network (RNN). The DNNs we tested (see Section~\ref{sec:implementation}) include two CNNs and one RNN. We provide a brief description of each architecture below and refer the interested readers to~\cite{Goodfellow-et-al-2016-Book} for more detailed descriptions.

\begin{figure}
    \centering
    \begin{minipage}{.5\textwidth}
        \centering
        \stackunder[3pt]{\includegraphics[width=0.95\columnwidth]{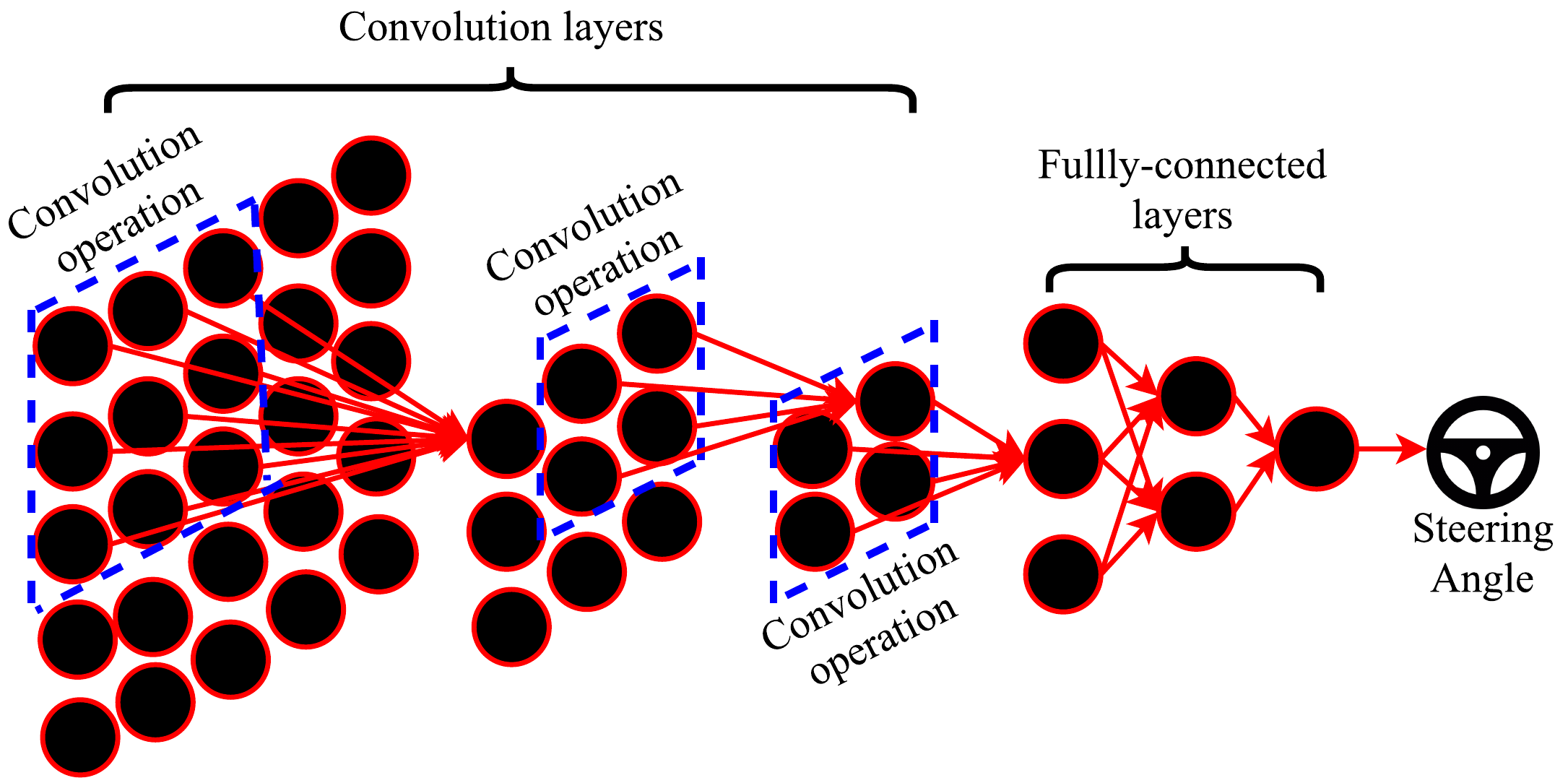}}{\scriptsize 3.1 A simplified CNN architecture}        
        \label{fig:cnn}
    \end{minipage}%
    \\
    \begin{minipage}{0.5\textwidth}
        \centering
        \stackunder[5pt]{\includegraphics[width=0.9\columnwidth]{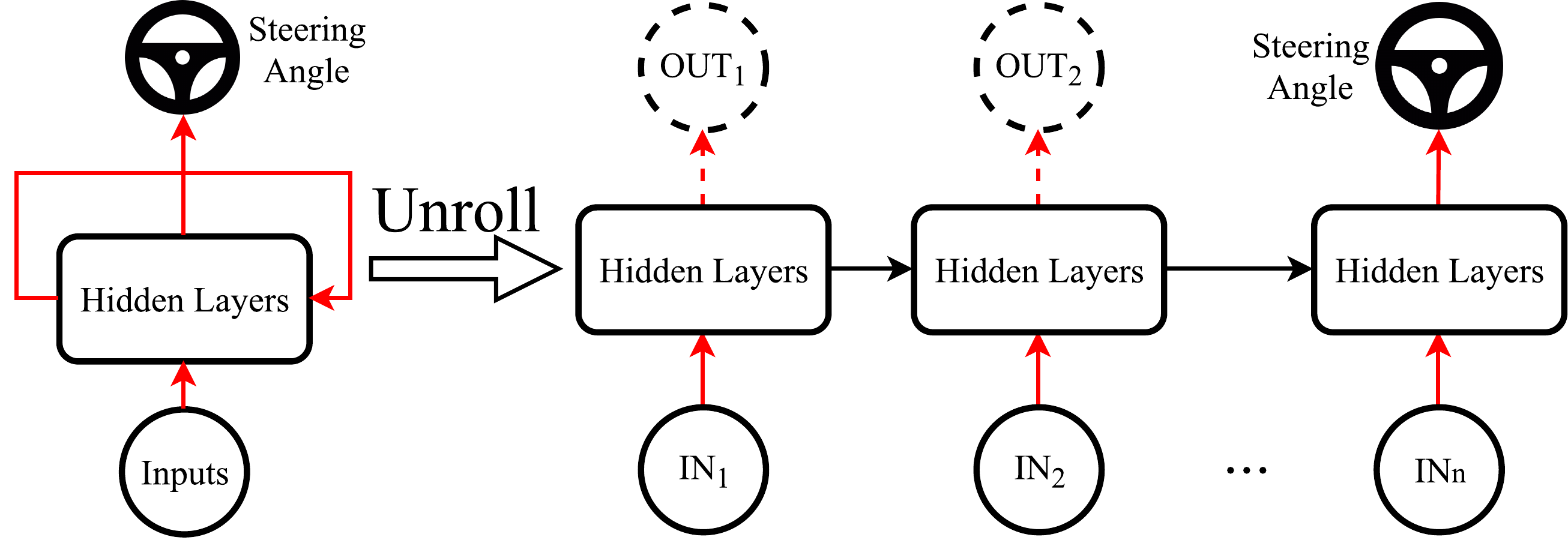}}
        {\scriptsize 3.2 A simplified RNN architecture}
        \label{fig:rnn}
    \end{minipage}
\caption{\small (Upper row) A simplified CNN architecture with a convolution kernel shown on the top-left part of the input image. The same filter (edges with same weights) is then moved across the entire input space, and the dot products are computed between the edge weights and the outputs of the connected neurons.  (Lower row) A simplified RNN architecture with loops in its hidden layers. The unrolled version on the right shows how the loop allows a sequence of inputs (\ie images) to be fed to the RNN and the steering angle is predicted based on all those images.}
\label{fig:cnn_rnn}
\vspace{-0.4cm}
\end{figure}

\noindent\textbf{CNN architecture.}
The most significant difference between a CNN and a fully connected DNN is the presence of a \textit{convolution layer}. The neurons in a convolution layer are connected only to some of the neurons in the next layer and multiple connections among different neurons share the same weight. The sets of connections sharing the same weights are essentially a convolution kernel~\cite{cnn} that applies the same convolution operation on the outputs of a set of neurons in the previous layer. Figure~\ref{fig:cnn_rnn} (upper row) illustrates the convolution operations for three convolution layers. This simplified architecture is similar to the ones used in practice~\cite{bojarski2016end}.

Convolution layers have two major benefits. First, they greatly reduce the number of trainable weights by allowing sharing of weights among multiple connections and thus significantly cut down the training time. Second, the application of convolution kernels is a natural fit for image recognition as it resembles the human visual system which extracts a layer-wise representation of visual input~\cite{cnn, krizhevsky2012imagenet}.

\noindent
\textbf{RNN architecture.}
RNNs, unlike CNNs, allow \textit{loops} in the network~\cite{rnn}. Specifically, the output of each layer is not only fed to the following layer but also flow back to the previous layer. Such arrangement allows the prediction output for previous inputs (e.g., previous frames in a video sequence) to be also considered in predicting current input. 
 Figure~\ref{fig:cnn_rnn} (lower row) illustrates a simplified version of the RNN architecture. 


Similar to other types of DNNs, RNNs also leverage gradient descent with back propagation for training. However, it is well known that the gradient, when propagated through multiple loops in an RNNs, may vanish to zero or explode to an extremely large value~\cite{hochreiter2001gradient} and therefore may lead to an inaccurate model.  \textit{Long short-term memory} (LSTM)~\cite{hochreiter1997long}, a popular subgroup of RNNs, is designed to solve this vanishing/exploding gradient problem. We encourage interested readers to refer to~\cite{hochreiter1997long} for more details.



\section{Methodology}
\label{sec:method}

To develop an automated testing methodology for DNN-driven 
autonomous cars we must answer the following questions. 
(i) How do we systematically explore the input-output spaces of an autonomous car DNN? 
(ii) How can we synthesize realistic inputs to automate such exploration? 
(iii) How can we optimize the exploration process? 
(iv) How do we automatically create a test oracle that can detect erroneous behaviors without detailed manual specifications?  
We briefly describe how \pname addresses each of these questions below. 


\subsection{Systematic Testing with Neuron Coverage}
\label{subsec:neuron_cov}
The input-output space (i.e., all possible combinations of inputs and outputs) of a complex system like an autonomous vehicle is too large for exhaustive exploration. Therefore, we must devise a systematic way of partitioning the space into different equivalence classes and try to cover all equivalence classes by picking one sample from each of them. In this paper, we leverage neuron coverage~\cite{pei2017deepxplore} as a mechanism for partitioning the input space based on the assumption that all inputs that have similar neuron coverage are part of the same equivalence class (i.e., the target DNN behaves similarly for these inputs).

Neuron coverage was originally proposed by Pei \etal for guided differential testing of multiple similar DNNs~\cite{pei2017deepxplore}. It is defined as the ratio of unique neurons that get activated for given input(s) and the total number of neurons in a DNN:
\begin{equation}\footnotesize
\label{eq:neuron_cov}
\begin{split}
    Neuron\ \ Coverage = \frac{|Activated\ \ Neurons|}{|Total\ \ Neurons|} 
\end{split}
\end{equation}
An individual neuron is considered activated if the neuron's output (scaled by the overall layer's outputs) is larger than a DNN-wide threshold. In this paper, we use $0.2$ as the neuron activation threshold for all our experiments.

Similar to the code-coverage-guided testing tools for traditional software, \pname tries to generate inputs that maximize neuron coverage of the test DNN. As each neuron's output affects the final output of a DNN, maximizing neuron coverage also increases output diversity. We empirically demonstrate this effect in Section~\ref{sec:result}.

Pei \etal defined neuron coverage only for CNNs~\cite{pei2017deepxplore}. We further generalize the definition to include RNNs. Neurons, depending on the type of the corresponding layer, may produce different types of output values (\ie single value and multiple values organized in a multidimensional array). We describe how we handle such cases in detail below.

For all neurons in fully-connected layers, we can directly compare their outputs against the neuron activation threshold as these neurons output a single scalar value. By contrast, neurons in convolutional layers output multidimensional feature maps as each neuron outputs the result of applying a convolutional kernel across the input space~\cite{hijazi2015using}. For example, the first layer in Figure~\ref{fig:cnn_rnn}.1 illustrates the application of one convolutional kernel (of size 3$\times$3) to the entire image (5$\times$5) that produces a feature map of size 3$\times$3 in the succeeding layer. In such cases, we compute the average of the output feature map to convert the multidimensional output of a neuron into a scalar and compare it to the neuron activation threshold.

For RNN/LSTM with loops, the intermediate neurons are unrolled to produce a sequence of outputs (Figure~\ref{fig:cnn_rnn}.2). We treat each neuron in the unrolled layers as a separate individual neuron for the purpose of neuron coverage computation. 

\subsection{Increasing Coverage with Synthetic Images}
Generating arbitrary inputs that maximize neuron coverage may not be very useful if the inputs are not likely to appear in the real-world even if these inputs potentially demonstrate buggy behaviors. Therefore, \pname focuses on generating realistic synthetic images by applying image transformations on seed images and mimic different real-world phenomena like camera lens distortions, object movements, different weather conditions, etc. To this end, we investigate nine different realistic image transformations (changing brightness, changing contrast, translation, scaling, horizontal shearing, rotation, blurring, fog effect, and rain effect). These transformations can be classified into three groups: linear, affine, and convolutional. Our experimental results, as described in Section~\ref{sec:result}, demonstrate that all of these transformations increase neuron coverage significantly for all of the tested DNNs. Below, we describe the details of the transformations.

Adjusting brightness and contrast are both linear transformations. The brightness of an image depends on how large the pixel values are for that image. An image's brightness can be adjusted by adding/subtracting a constant parameter $\beta$ to each pixel's current value. Contrast represents the difference in brightness between different pixels in an image. One can adjust an image's contrast by multiplying each pixel's value by a constant parameter $\alpha$.

\begin{table}[h!tbp]
  \centering
  \setlength{\tabcolsep}{4pt}
    \scriptsize
    \renewcommand{\arraystretch}{.9}
  \caption{\small Different affine transformation matrices}
    \begin{tabular}{l|l|l|l}
    
\textbf{Affine Transform} & \textbf{Example} & \textbf{Transformation Matrix} & \textbf{Parameters} \\
\toprule
\textbf{Translation} &\includegraphics[width=0.04\textwidth, valign=c]{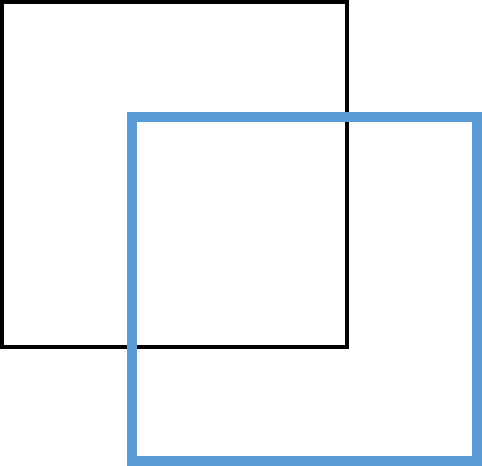} & 
\( \begin{bmatrix}1 & 0 & t_x \\0 & 1 & t_y\end{bmatrix} \)  & \makecell{$t_x$: displacement along x axis \\$t_y$: displacement along y axis}\\

\midrule
\textbf{Scale} &\includegraphics[width=0.04\textwidth, valign=c]{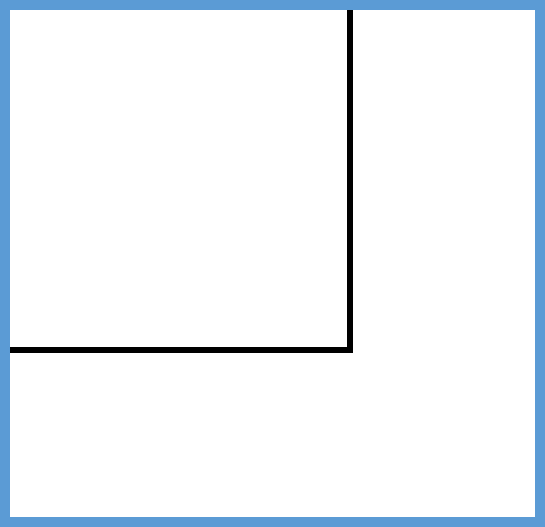}& \( \begin{bmatrix}s_x & 0 & 0 \\0 & s_y & 0\end{bmatrix} \)& \makecell{$s_x$: scale factor along x axis \\$s_y$: scale factor along y axis}\\
\midrule
\textbf{Shear} & \includegraphics[width=0.04\textwidth, valign=c]{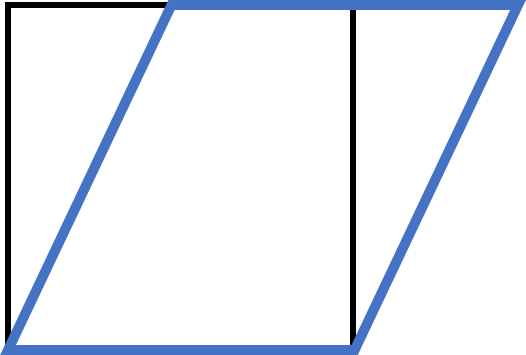}& \( \begin{bmatrix}1 & s_x & 0 \\s_y & 1 & 0\end{bmatrix} \) & \makecell{$s_x$: shear factor along x axis \\$s_y$: shear factor along y axis}\\
\midrule
\textbf{Rotation} & \includegraphics[width=0.04\textwidth, valign=c]{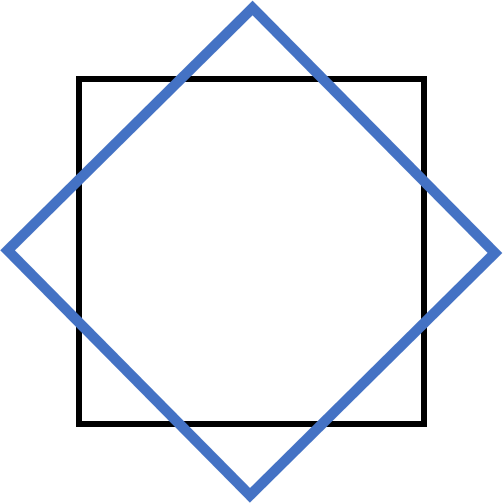}& \( \begin{bmatrix}\cos q & -\sin q & 0 \\ \sin q & \cos q & 0\end{bmatrix} \)& $q$: the angle of rotation\\
    \bottomrule
    \end{tabular}%
  \label{tab:affine}%
  \vspace{-0.4cm}
\end{table}%

Translation, scaling, horizontal shearing, and rotation are all different types of affine transformations. An affine transformation is a linear mapping between two images that preserves points, straight lines, and planes~\cite{affinetransform1}.
Affine transforms are often used in image processing to fix distortions resulting from camera angle variations. In this paper, we leverage affine transformations for the inverse case, i.e., to simulate different real-world camera perspectives or movements of objects and check how robust the self-driving DNNs are to those changes. 

An affine transformation is usually represented by a $2\times3$ transformation matrix $M$~\cite{affinetransform2}. One can apply an affine transformation to a 2D image matrix $I$ by simply computing the dot product of $I$ and $M$, the corresponding transformation matrix. We list the transformation matrices for the four types of affine transformations (translation, scale, shear, and rotation) used in this paper in Table~\ref{tab:affine}.

Blurring and adding fog/rain effects are all convolutional transformations, i.e., they perform the convolution operation on the input pixels with different transform-specific kernels. A convolution operation adds (weighted by the kernel) each pixel of the input image to its local neighbors. We use four different types of blurring filters: averaging, Gaussian, median, and bilateral~\cite{itseez2014theopencv}. We compose multiple filters provided by Adobe Photoshop on the input images to simulate realistic fog and rain effects~\cite{raineffect, fogeffect}.
 
\subsection{Combining Transformations to Increase Coverage}
As the individual image transformations increase neuron coverage, 
one obvious question is whether they can be combined to further increase the neuron coverage. Our results demonstrate that different image transformations tend to activate different neurons, i.e., they can be stacked together to further increase neuron coverage. However, the state space of all possible combinations of different transformations is too large to explore exhaustively. We provide a neuron-coverage-guided greedy search technique for efficiently finding combinations of image transformations that result in higher coverage (see Algorithm~\ref{algo:guided}).

\vspace{-0.4cm}
\begin{algorithm}[!htb]
\scriptsize
	\SetAlgoLined
	\caption{\textbf{\small Greedy search for combining image tranformations to increase neuron coverage}}
	\label{algo:guided}
	\SetKwInOut{Input}{Input}
	\SetKwInOut{Output}{Output}
	\SetKwInOut{Variable}{Variable}
	\Input{Transformations T, Seed images I}
	\Output{Synthetically generated test images}
	\Variable{S: stack for storing newly generated images \\
		  Tqueue: transformation queue \\
        }
    \hrulefill \\
	
	Push all seed imgs $\in$ I to Stack S \\
	genTests = $\phi$\\
	\While{$S$ is not empty} {
		img = S.pop() \\
		Tqueue = $\phi$ \\
		numFailedTries = 0 \\
		\While{numFailedTries $\le$ maxFailedTries}{
			\eIf {Tqueue is not empty}{
				T1 = Tqueue.dequeue()\\
			}
			{ 
				Randomly pick transformation T1 from T \\
			}
			Randomly pick parameter P1 for T1\\
			    
			Randomly pick transformation T2 from T \\
			Randomly pick parameter P2 for T2\\
			
			newImage = ApplyTransforms(image, T1, P1, T2, P2)\\
			\eIf{covInc(newimage)}{
				Tqueue.enqueue(T1) \\
			        Tqueue.enqueue(T2) \\
				UpdateCoverage()  \\
			        genTest = genTests $\cup$ newimage
			        S.push(newImage) \\
			}
			{
				numFailedTries = numFailedTries + 1 \\
			}
		}	
	}
	return genTests \\
\end{algorithm}
\vspace{-0.4cm}

The algorithm takes a set of seed images $I$, a list of transformations T and their corresponding parameters as input. The key idea behind the algorithm is to keep track of the transformations that successfully increase neuron coverage for a given image and prioritize them while generating more synthetic images from the given image. This process is repeated in a depth-first manner to all images.

\subsection{Creating a Test Oracle with Metamorphic Relations}
\label{subsec:mr}

One of the major challenges in testing a complex DNN-based system like an autonomous vehicle is creating the system's specifications manually, against which the system's behavior can be checked. 
It is challenging to create detailed specifications for such a system as it essentially involves recreating the logic of a human driver.
To avoid this issue, we leverage metamorphic relations~\cite{chen1998metamorphic} between the car behaviors across different synthetic images. The key insight is that even though it is hard to specify the correct behavior of a self-driving car for every transformed image, one can define relationships between the car's behaviors across certain types of transformations. For example, the autonomous car's steering angle should not change significantly for the same image under any lighting/weather conditions, blurring, or any affine transformations with small parameter values. Thus, if a DNN model infers a steering angle $\theta_o$ for an input seed image \iorg~ and    
a steering angle $\theta_t$ for a new synthetic image \itrans, which is generated by applying the  transformation $t$ on \iorg, one may define a simple metamorphic relation where $\theta_o$  and $\theta_t$ are identical. 

However, there is usually no single correct steering angle for a given image, i.e., a car can safely tolerate small variations. Therefore, there is a trade-off between defining the metamorphic relations very tightly, like the one described above (may result in a large number of false positives) and making the relations more permissive (may lead to many false negatives). In this paper, we strike a balance between these two extremes by using the metamorphic relations defined below.

To minimize false positives, we relax our metamorphic relations and allow variations within the error ranges of the original input images. We observe that the set of outputs predicted by a DNN model for the original images, say \{$\theta_{o1}, \theta_{o2}, .... ,\theta_{on}$\},  in practice, result in a small but non-trivial number of errors \wrt their respective manual labels (\{$\hat\theta_{1}, \hat\theta_{2}, .... ,\hat\theta_{n}$\}). Such errors are usually measured using Mean Squared Error (MSE), where $MSE_{orig} = \frac{1}{n}\sum_{i=1}^n(\hat\theta_{i}-\theta_{oi})^2$. Leveraging this property, we redefine a new metamorphic relation as:

\begin{equation}
\label{eq:mr}
(\hat\theta_{i} - \theta_{ti})^2 \le \lambda\ MSE_{orig}
\end{equation}

The above equation assumes that the errors produced by a model for the transformed images as input should be within a range of $\lambda$ times the MSE produced by the original image set. Here, $\lambda$ is a configurable parameter that allows us to strike a balance between the false positives and false negatives. 

\section{Implementation}
\label{sec:implementation}

\noindent
{\bf Autonomous driving DNNs.}
We evaluate our techniques on three DNN models that won top positions in the Udacity self-driving challenge~\cite{challenge2}: Rambo~\cite{rambo} (2$^{nd}$ rank), Chauffeur~\cite{chauffeur} (3$^{rd}$ rank), and Epoch~\cite{epoch} (6$^{th}$ rank). We choose these three models as their implementations are based on the Keras framework~\cite{chollet2015keras} that our current prototype of \pname supports. 
The details of the DNN models and dataset are summarized in Table~\ref{tab:dnn_steering_angle}.




As shown in the right figure of Table~\ref{tab:dnn_steering_angle}, the steering angle is defined as the rotation degree between the heading direction of the vehicle (the vertical line) and the heading directions of the steering wheel axles (\ie, usually front wheels). The negative steering angle indicates turning left while the positive values indicate turning left.
The maximum steering angle of a car varies based on the hardware of different cars. 
The Udacity self-driving challenge dataset used in this paper has a maximum steering angle of +/- 25 degree~\cite{challenge2}.
The steering angle is then scaled by 1/25 so that the prediction should fall between -1 and 1. 

\begin{table}[htbp]
\begin{minipage}[c]{0.55\columnwidth}
\centering
\setlength{\tabcolsep}{1pt}
\scriptsize
\renewcommand{\arraystretch}{.8}
\begin{threeparttable}
    \centering
    \begin{tabular}{lllll}
        &  & \textbf{No. of} & \textbf{Reported} & \textbf{Our} \\
\textbf{Model} & \textbf{Sub-Model} & \textbf{Neurons}& \textbf{MSE} & \textbf{MSE} \\
\toprule
\multirow{2}[0]{*}{\textbf{\chauf}} & CNN & 1427 & \multirow{2}[0]{*}{0.06} & \multirow{2}[0]{*}{0.06}\\
          & LSTM & 513 & &  \\
          
\midrule

\multirow{3}[0]{*}{\textbf{\rambo}} & S1(CNN)  & 1625 & \multirow{3}[0]{*}{0.06} & \multirow{3}[0]{*}{0.05}\\
          
          & {S2}(CNN)   & 3801 & & \\
          & {S3}(CNN)   & 13473 & & \\
          
\midrule
    \textbf{\epoch} &  CNN    & 2500 & 0.08 & 0.10 \\
    \bottomrule
    
    \end{tabular}%
    \begin{tablenotes}
      \scriptsize
      \item[\wu]  dataset HMB\_3.bag~\cite{dataset}
    \end{tablenotes}
\end{threeparttable}
\end{minipage}
\hfill
\begin{minipage}[c]{0.43\columnwidth}
\centering
\includegraphics[width=3.5cm]{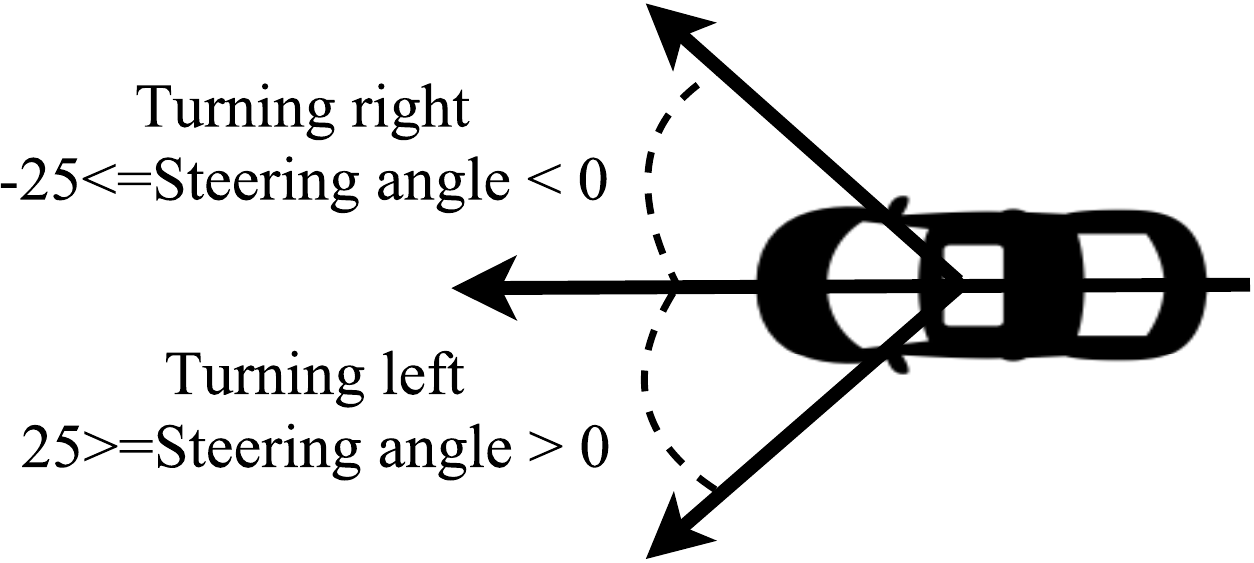}
\end{minipage}
\vspace{.3cm}
	\caption{\small (Left) Details of DNNs used to evaluate \pname.\wu (Right) The outputs of the DNNs are the steering angles for a self-driving car heading forward. The Udacity self-driving car has a maximum steering angle of +/- 25 degree.}
\label{tab:dnn_steering_angle}
\vspace{-0.25 in}
\end{table}


\textbf{Rambo} model consists of three CNNs whose outputs are merged using a final layer~\cite{rambo}. Two of the CNNs are inspired by NVIDIA's self-driving car architecture~\cite{bojarski2016end}, and the third CNN is based on comma.ai's steering model~\cite{comma.ai}. As opposed to other models that take individual images as input, \rambo takes the differences among three consecutive images as input. The model uses Keras~\cite{chollet2015keras} and Theano~\cite{2016arXiv160502688short} frameworks.

\textbf{Chauffeur} model includes one CNN model for extracting features from the image and one LSTM model for predicting steering angle~\cite{chauffeur}. The input of the CNN model is an image while the input of the LSTM model is the concatenation of 100 features extracted by the CNN model from previous 100 consecutive images. Chauffeur uses Keras~\cite{chollet2015keras} and Tensorflow~\cite{abadi2016tensorflow} frameworks.

\textbf{Epoch} model uses a single CNN. As the pre-trained model for \epoch is not publicly available, we train the model using the instructions provided by the authors~\cite{epoch}. We used the  CH2\_002 dataset~\cite{dataset} from the Udacity self-driving Challenge for training the epoch model. \epoch, similar to Chauffeur, uses Keras and Tensorflow frameworks.

\noindent
\textbf{Image transformations.}
In the experiments for RQ2 and RQ3, we leverage seven different types of simple image transformations: translation, scaling, horizontal shearing, rotation, contrast adjustment, brightness adjustment, and blurring. We use OpenCV to implement these transformations~\cite{itseez2015opencv}. For RQ2  and RQ3 described in Section~\ref{sec:result}, we use 10 parameters for each transformation as shown in Table~\ref{tab:setup}. 


\begin{table}[htbp]
\setlength{\tabcolsep}{0.5pt}
  \centering
  \setlength{\tabcolsep}{4pt}
    \scriptsize
    \renewcommand{\arraystretch}{.8}
  \caption{\small Transformations and parameters used by \pname for generating synthetic images.}

    \begin{tabular}{lc|lc}
	    & \textbf{Transformations} & \textbf{Parameters}  &\textbf{Parameter ranges} \\
\toprule
 &    \textbf{Translation} &$(t_x,t_y)$ &\makecell{$(10,10)$ to $(100,100)$\\ step $(10,10)$}\\ \midrule
&	 \textbf{Scale} &$(s_x,s_y)$ &\makecell{$(1.5,1.5)$ to $(6,6)$\\ step $(0.5,0.5)$} \\ \midrule
 &    \textbf{Shear} &$(s_x,s_y)$ &\makecell{$(-1.0,0)$ to $(-0.1,0)$\\ step $(0.1,0)$} \\ \midrule
  &   \textbf{Rotation} &$q$ (degree) &\makecell{$3$ to $30$ with step $3$} \\ \midrule
   &  \textbf{Contrast} &$\alpha$ (gain) &\makecell{$1.2$ to $3.0$ with step $0.2$} \\ \midrule
    & \textbf{Brightness} &$\beta$ (bias) &\makecell{$10$ to $100$ with step $10$} \\ \midrule
    & \textbf{Averaging} & kernel size & $3 \times 3$, $4 \times 4$, $5 \times 5$, $6 \times 6$ \\ \cmidrule{2-4}
    & \textbf{Gaussian} & kernel size & $3 \times 3$, $5 \times 5$, $7 \times 7$ , $3 \times 3$ \\ \cmidrule{2-4}
	    \multicolumn{1}{c}{}{\textbf{Blur}}   & \textbf{Median} &aperture linear size & 3, 5 \\ \cmidrule{2-4}
     & \textbf{Bilateral Filter} &diameter, sigmaColor, sigmaSpace & 9, 75, 75 \\
    \bottomrule
    
    \end{tabular}%
  \label{tab:setup}%
  \vspace{-0.4cm}
\end{table}%

\section{Results}
\label{sec:result}




As DNN-based models are fundamentally different than traditional software, first, we check whether neuron coverage is a good metric to capture functional diversity of DNNs.  In particular, we investigate whether neuron coverage changes with different input-output pairs of an autonomous car. An individual neuron's output goes through a sequence of linear and nonlinear operations before contributing to the final outputs of a DNN. Therefore, it is not very clear how much (if at all) individual neuron's activation will change the final output. We address this in our first research question.


\begin{table}[htbp]
  \centering
  \setlength{\tabcolsep}{5pt}
    \scriptsize
    \renewcommand{\arraystretch}{.8}
  \caption{\small Relation between neuron coverage and test output}
    \begin{tabular}{l|l|r|ll}
 &  & \multicolumn{1}{c|}{\textbf{Steering}} & \multicolumn{2}{c}{\textbf{Steering}} \\
\textbf{Model} & \textbf{Sub-Model} & \multicolumn{1}{c|}{\textbf{Angle}} & \multicolumn{2}{c}{\textbf{Direction}} \\
\toprule
      &       & Spearman  & Wilcoxon  & Effect size  \\
      &       & Correlation & Test & (Cohen's d) \\
    \midrule
    \textbf{\chauf} & Overall & -0.10 (***) & \positive $>$ \negative (***) & negligible \\
        \multirow{2}[0]{*}{} & {CNN} & 0.28 (***) & \positive $<$ \negative (***) & negligible \\
          & LSTM & -0.10 (***)& \positive $>$ \negative (***) & negligible \\
          \midrule
    \textbf{\rambo} & Overall & -0.11 (***) & \positive $<$ \negative (***) & negligible \\
          & {S1} & -0.19 (***) & \positive $<$ \negative (***) & large \\
          & {S2} & 0.10 (***) & not significant & negligible \\
          & {S3} & -0.11 (***) & not significant & negligible \\
          \midrule
    \textbf{\epoch} & N/A & 0.78 (***) & \positive $<$ \negative (***) & small \\
    \bottomrule
    \end{tabular}%
	\\ *** indicates statistical significance with p-value $<$ $2.2*10^{-16}$
  \label{tab:rq1}%
\end{table}%

\RQ{1}{Do different input-output pairs result in different neuron coverage?}


For each input image we measure the neuron coverage (see Equation~\ref{eq:neuron_cov} in Section~\ref{subsec:neuron_cov}) of the underlying models and the corresponding output. As discussed in Section~\ref{sec:implementation}, corresponding to an input image, each model outputs a steering direction (left (+ve) / right (-ve)) and a steering angle as shown in Table~\ref{tab:dnn_steering_angle} (right). We analyze the neuron coverage for both of these outputs separately.

\textbf{Steering angle}. As steering angle is a continuous variable, we check Spearman rank correlation~\cite{spearman1904proof}  between neuron coverage and steering angle. This is a non-parametric measure to compute monotonic association between the two variables~\cite{hauke2011comparison}. Correlation with positive statistical significance suggests that the steering angle increases with increasing neuron coverage and vice versa. 
Table~\ref{tab:rq1} shows that Spearman correlations for all the models are statistically significant\textemdash while \chauf and \rambo models show an overall negative association, \epoch model shows a strong positive correlation. This result indicates that the neuron coverage changes with the changes in output steering angles, \ie different neurons get activated for different outputs. Thus, in this setting, neuron coverage can be a good approximation for estimating the diversity of input-output pairs. Moreover, our finding that monotonic correlations between neuron coverage and steering angle also corroborate Goodfellow et al.'s hypothesis that, in practice, DNNs are often highly linear~\cite{goodfellow2014explaining}.

\textbf{Steering direction}. 
To measure the association between neuron coverage and steering direction, we check whether the coverage varies between right and left steering direction. We use the Wilcoxon nonparametric test as the steering direction can only have two values (left and right). Our results confirm that neuron coverage varies with steering direction with statistical significance (p $<$ $2.2*10^{-16}$) for all the three overall models. Interestingly, for \rambo, only the Rambo-S1 sub-model shows statistically significant correlation but not Rambo-S2 and Rambo-S3. These results suggest that, unlike steering angle, some sub-models are more responsible than other for changing steering direction.  

Overall, these results show that neuron coverage altogether varies significantly for different input-output pairs. Thus, a neuron-coverage-directed (NDG) testing strategy can help in finding corner cases.

\RS{1}{Neuron coverage is correlated with input-output diversity and can be used to systematic test generation.}

\begin{figure*}[!htpb]
\centering
\stackunder[5pt]{
\includegraphics[width=0.85\columnwidth]{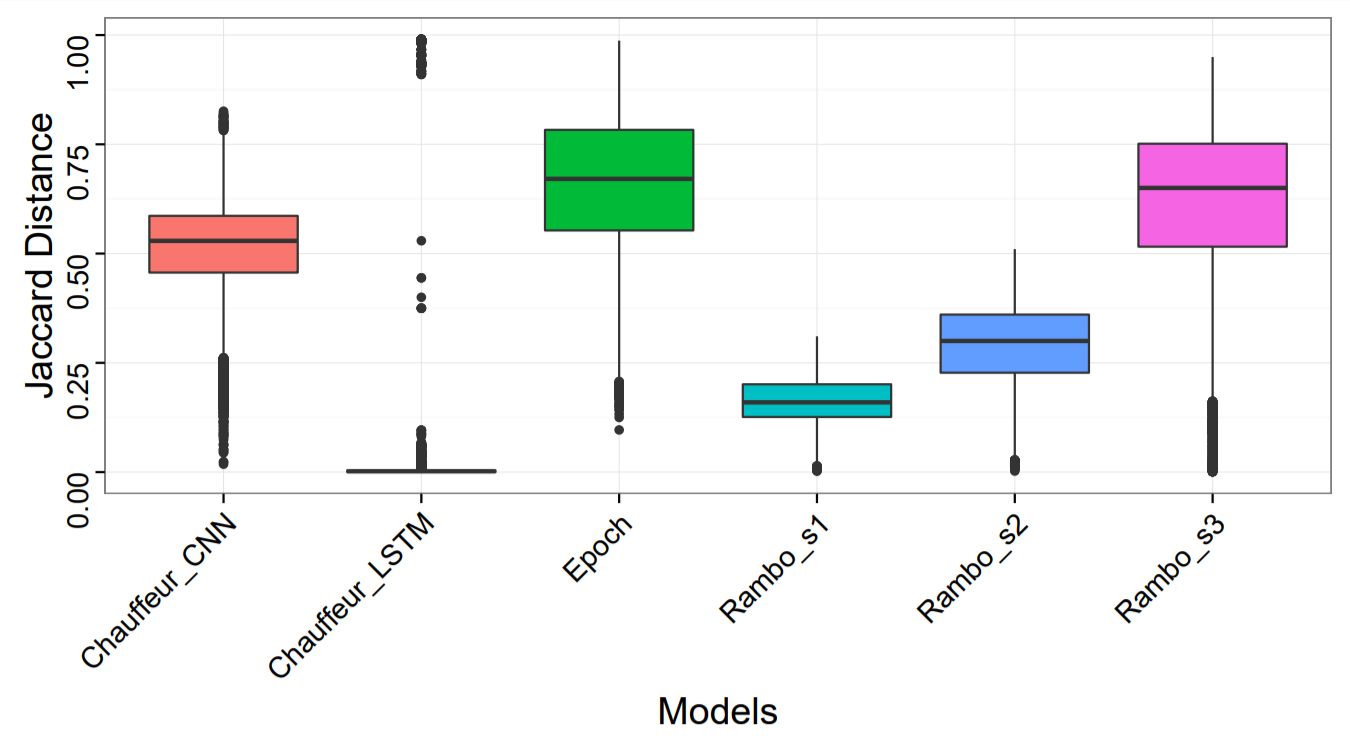}}{\scriptsize 4.1 Difference in neuron coverage caused by different image transformations}
\qquad
\qquad
 \stackunder[5pt]{
\includegraphics[width=0.85\columnwidth]{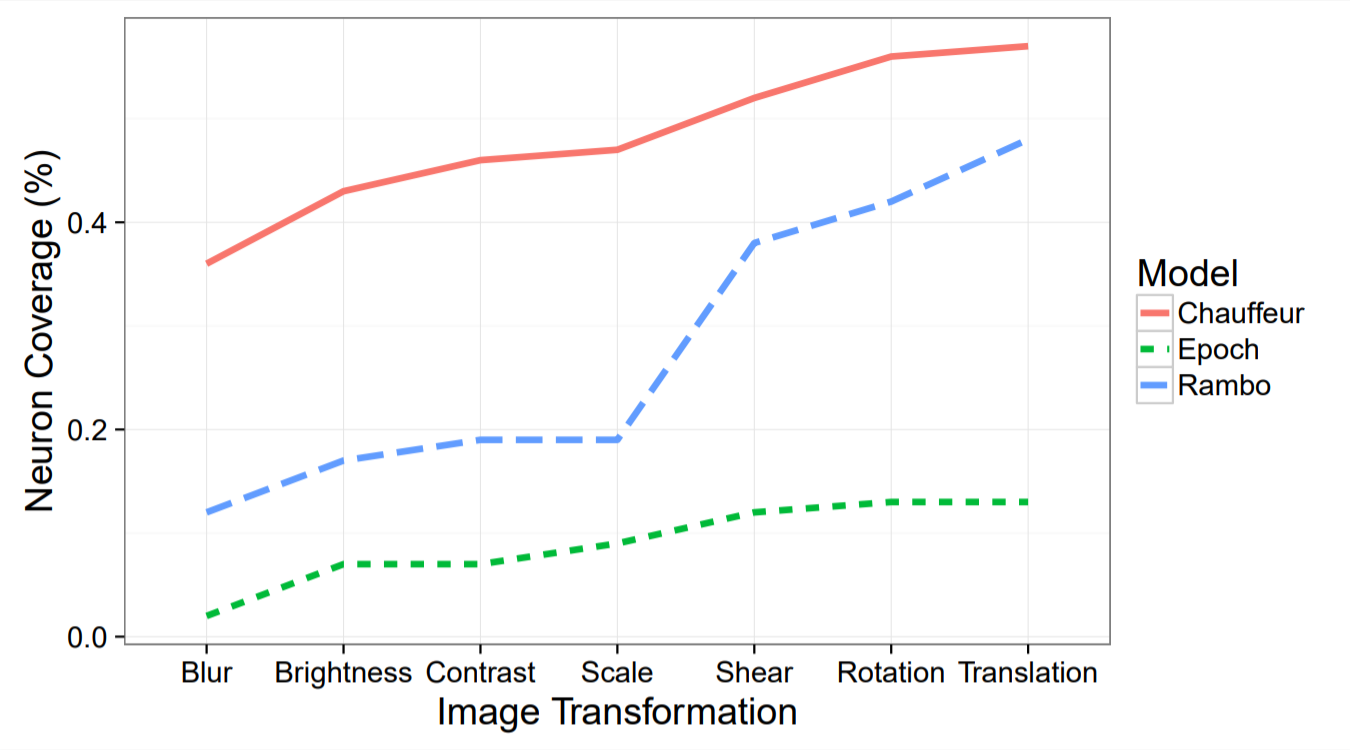}}{\scriptsize 4.2 Average cumulative neuron coverage per input image}
\caption{\small Different image transformations activate significantly different neurons. In the top figure the median Jaccard distances for Chauffeur-CNN, Chauffeur-LSTM, Epoch, Rambo-S1, Rambo-S2, and Rambo-S3 models are 0.53, 0.002, 0.67, 0.12, 0.17, 0.30, and 0.65.}
\label{fig:rq2}
\vspace{-0.2 cm}
\end{figure*}

Next, we investigate whether synthetic images generated by applying different realistic image transformations (as described in Table~\ref{tab:affine}) on existing input images can activate different neurons. Thus, we check: \\

\RQ{2}{Do different realistic image transformations activate different neurons?}

We randomly pick 1,000 input images from the test set and transform each of them by using seven different transformations: blur, brightness, contrast, rotation,  scale,  shear, and translation. We also vary the parameters of each transformation and generate a total of 70,000 new synthetic images.  We run all models with these synthetic images as input and record the neurons activated by each input. 

We then compare the neurons activated by different synthetic images generated from the same image. 
Let us assume that two transformations $T_1$ and $T_2$, when applied to an original image $I$, activate two sets of neurons $N_1$ and $N_1$ respectively. We measure the dissimilarities between $N_1$ and $N_2$ by measuring their Jaccard distance: $1 - \frac{|N_{1} \cap N_{2}|}{|N_{1} \cup N_{2}|}$.  

Figure~\ref{fig:rq2}.1 shows the result for all possible pair of transformations (\eg blur vs.~rotation, rotation vs.~transformation, \etc) for different models. These results indicate that for all models, except \lstm, different transformations activate different neurons. As discussed in Section~\ref{subsec:dl_background}, LSTM is a particular type of RNN architecture that keeps states from previous inputs and hence increasing the neuron coverage of LSTM models with single transformations is much harder than other models. In this paper, we do not explore this problem any further and leave it as an interesting future work. 

We further check how much a single transformation contributes in increasing the neuron coverage \wrt all other transformations for a given seed image. Thus, if an original image $I$ undergoes seven discrete transformations: $T_1, T_2,...T_7$, we compute the total number of neurons activated by $T_1$, $T_1 \cup T_2$, ..., $\bigcup\limits_{i=1}^7 T_i$.  Figure~\ref{fig:rq2}.2 shows the cumulative effect of all the transformations on average neuron coverage per seed image. We see that the cumulative coverage increases with increasing number of transformations for all the models. In other words, all the transformations are contributing towards the overall neuron coverage. 

\begin{figure}[!htpb]
\centering
\includegraphics[width=\columnwidth]{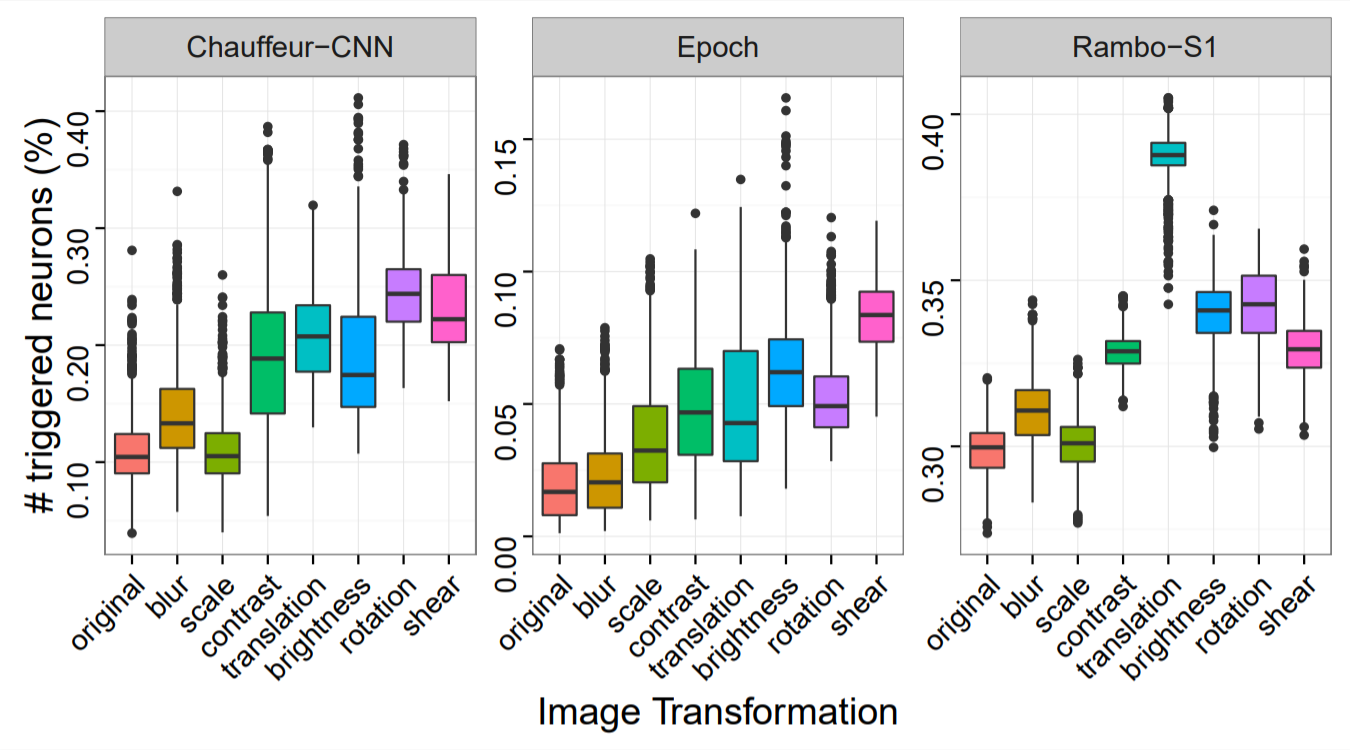}

{\scriptsize
\begin{tabular}{lrrr}
\multicolumn{4}{c}{\textbf{\small Median Increase in Neuron Coverage}} \\
\toprule
\toprule
\textbf{Transformation} & \textbf{\chauf} & \textbf{\epoch} & \textbf{\rambo} \\
 & (\textbf{CNN},\textbf{LSTM}) &  & (\textbf{S1},\textbf{S2},\textbf{S3})    \\
\toprule
Scale & (1.0,0.0) & 39.0** & (2.0*,5.0*,32.0)   \\
      & (0.67\%,0\%) & 93\% & (0.41\%,1\%,4\%)   \\
\midrule
Brightness & (100.0**,1.0) & 113.0** & (67.0**,104.0**,585.0*)   \\
           & (67\%,0.2\%)  & 269\%   & (14\%,24\%,66\%)   \\
\midrule
Contrast & (120.0**,1.0*) & 75.0** & (47.0**,100.0**,159.0)  \\
         & (80\%,0.2\%)   & 179\% & (10\%,23\%,18\%)   \\
\midrule
Blur & (41.0**,0.0) & 9.0*  & (18.0**,23.0**,269.5*)   \\
     & (28\%,0\%)   & 21\%  & (4\%,5\%,31\%)  \\
\midrule
Rotation & (199.0**,2.0*) & 81.0** & (70.0**,13.0**,786.5*)   \\
         & (134\%,0.39\%) & 193\%  & (14\%,3\%,89\%)   \\
\midrule
Translation & (147.0**,1.0*) & 65.0** & (143.0**,167.0**,2315.5**)   \\
            & (99\%,0.2\%)   & 155\%  & (29\%,38\%,263\%)  \\
\midrule
Shear & (168.0**,1.0*) & 167.0** & (48.0**,132.0**,1472.0**)   \\
      & (113\%,0.2\%)  & 398\%   & (10\%,30\%,167\%)   \\
\bottomrule
\end{tabular}%
\vspace{0.1cm}

{All numbers are statistically significant; \\ 
Numbers with * and ** have small and large Cohen's D effect.}

}

\caption{\textbf{\small Neuron coverage per seed image for individual image transformations \wrt baseline.}}
\label{fig:rq3}
\end{figure}

We also compute the percentage change in neuron coverage per image transformation ($N_{T}$) \wrt to the corresponding seed image ($N_{O}$) as: ($N_{T}$-$N_{O}$)/$N_{O}$. Figure~\ref{fig:rq3} shows the result. For all the studied models, the transformed images increase the neuron coverage significantly\textemdash Wilcoxon nonparametric test confirms the statistical significance. These results also show that different image transformations increase neuron coverage at different rates.

\RS{2}{Different image transformations tend to activate different sets of neurons.}

Next, we mutate the seed images with different combinations of transformations (see Section~\ref{sec:method}). 
Since different image transformations activate different set of neurons, here we try to increase the neuron coverage by these transformed image inputs. To this end, we question: 

\RQ{3}{Can neuron coverage be further increased by combining different image transformations?}

We perform this experiment by measuring neuron coverage in two different settings: (i) applying a set of transformations and (ii) combining transformations using \textit{coverage-guided} search. 

{i) \textit{Cumulative Transformations.}
Since different seed images activate a different set of neurons (see RQ1), 
multiple seed images collectively achieve higher neuron coverage than a single one. 
Hence, we check whether the transformed images can still increase the neuron coverage collectively \wrt the cumulative baseline coverage of a set of seed images. In particular, we generate a total of 7,000 images from 100 seed images by applying 7 transformations and varying 10 parameters on 100 seed images. 
This results in a total of 7,000 test images.
We then compare the cumulative neuron coverage of these synthetic images \wrt the baseline, which use the same 100 seed images for fair comparison. 
Table~\ref{tab:rq3_2} shows the result. Across all the models (except Rambo-S3), the cumulative coverage increased significantly. Since the \ru baseline already achieved 97\% coverage, the transformed images only increase the coverage by $(13,080 - 13,008)/13,008 = 0.55\%$.

\setlength{\tabcolsep}{2pt}
\begin{table}[htbp]
\centering
\scriptsize
\renewcommand{\arraystretch}{.8}
\caption{\small Neuron coverage achieved by cumulative and guided transformations applied to 100 seed images.}
\begin{tabular}{lrrr|rr}
 &  & \multicolumn{1}{c}{\textbf{Cumulative}} & \textbf{Guided} & \multicolumn{2}{c}{\textbf{\% increase of guided \wrt}} \\
\textbf{Model} & \textbf{Baseline} & \multicolumn{1}{c}{\textbf{Transformation}} & \textbf{Generation} & \textbf{Baseline} & \textbf{Cumulative}\\
    \toprule
    {\cnn}   & 658 (46\%)  & 1,065 (75\%)  & 1,250 (88\%) &  90\% & 17\% \\
    {\epoch} & 621 (25\%)  & 1034 (41\%)  & 1,266 (51\%) & 104\% & 22\%\\
    {\rs}    & 710 (44\%)  & 929  (57\%)  & 1,043 (64\%) & 47\% & 12\% \\
    {\rt}    & 1,146 (30\%) & 2,210 (58\%)  & 2,676 (70\%) & 134\% & 21\%\\
    {\ru}    & 13,008 (97\%) & 13,080 (97\%) & 13,150 (98\%) & 1.1\% & 0.5\%\\

    \bottomrule
    \end{tabular}%
  \label{tab:rq3_2}%
\vspace{-0.4cm}
\end{table}%

{ii) \textit{Guided Transformations.}
Finally, we check whether we can further increase the cumulative neuron coverage by using the coverage-guided search technique described in Algorithm~\ref{algo:guided}. We generate 254, 221, and 864 images from 100 seed images for \cnn, \epoch, and \rambo models respectively and measure their collective neuron coverage. As shown in Table~\ref{tab:rq3_2}, the guided transformations collectively achieve 88\%, 51\%, 64\%, 70\%, and 98\% of total neurons for models \cnn, \epoch, \rs, \rt, and \ru respectively, thus increasing the coverage up to 17\% 22\%, 12\%,  21\%, and 0.5\% \wrt the unguided approach. This method also significantly achieves higher neuron coverage \wrt baseline cumulative coverage.

\RS{3}{By systematically combining different image transformations, 
neuron coverage can be improved by around 100\% w.r.t. the coverage achieved by  the original seed images.}

Next we check whether the synthetic images can trigger any erroneous behavior in the autonomous car DNNs and if we can detect those behaviors using metamorphic relations as described in Section~\ref{subsec:mr}. This leads to the following research question:

\RQ{4}{Can we automatically detect erroneous behaviors using metamorphic relations?}

\begin{figure}[!htpb]
\centering
\includegraphics[width=0.8\columnwidth]{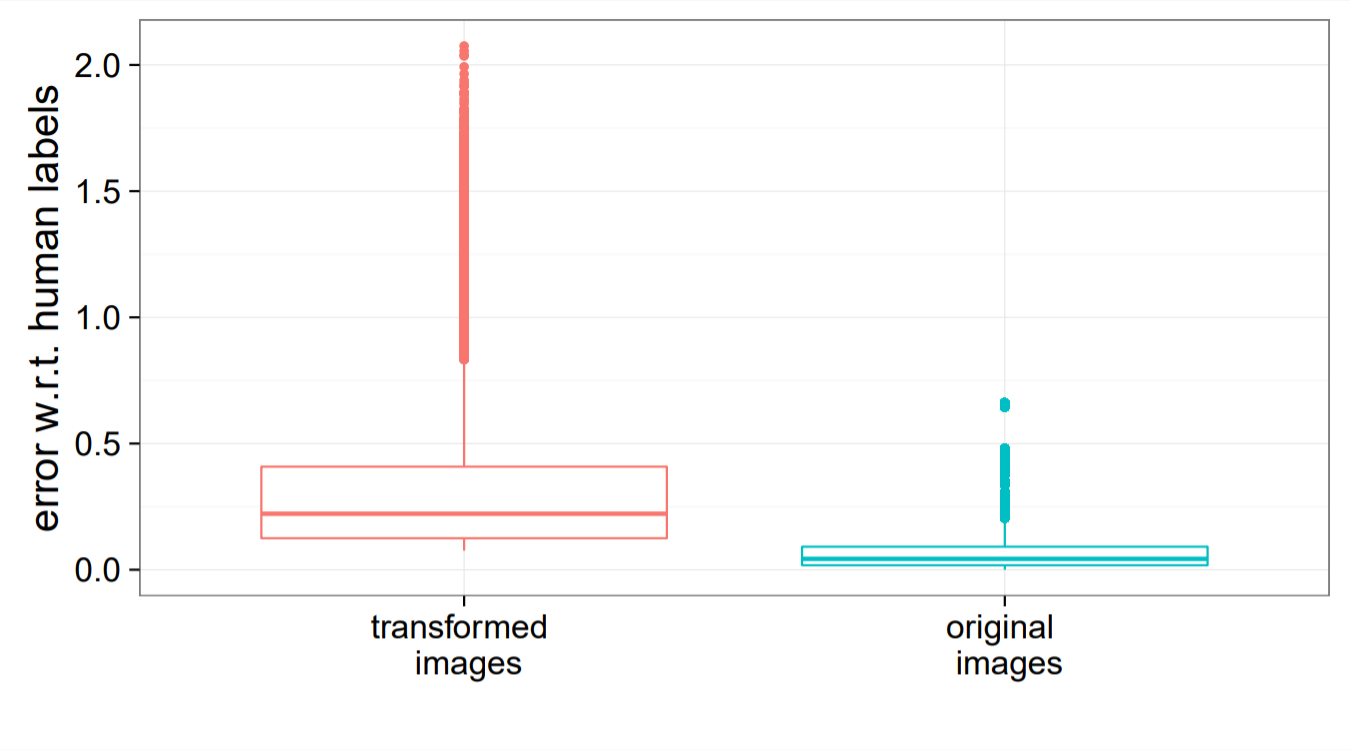}
\caption{\textbf{\small Deviations from the human labels for images that violate the metamorphic relation (see Equation~\ref{eq:mr}) is higher compared to the deviations for original images. Thus, these synthetic images have a high chance to show erroneous behaviors.}}
\label{fig:rq4}
\end{figure}

Here we focus on the transformed images whose outputs violate the metamorphic relation defined in Equation~\ref{eq:mr}. We call these images \isusp~and their corresponding original images as \iorig. We compare the deviation between the outputs of \isusp~and \iorig~ \wrt the corresponding human labels, as shown in Figure~\ref{fig:rq4}. The deviations produced for \isusp~are much larger than \iorig~ (also confirmed by Wilcoxon test for statistical significance). In fact, mean squared error (MSE) for \isusp~  is $0.41$, while the MSE of the corresponding \iorig~ is $0.035$. Such differences also exist for other synthetic images that are generated by composite transformations including rain, fog, and those generated during the coverage-guided search. Thus, overall \isusp~ has a higher potential to show buggy behavior.

\begin{figure*}



\stackunder[4pt]{
\includegraphics[width=0.09\textwidth, height=0.09\textwidth]{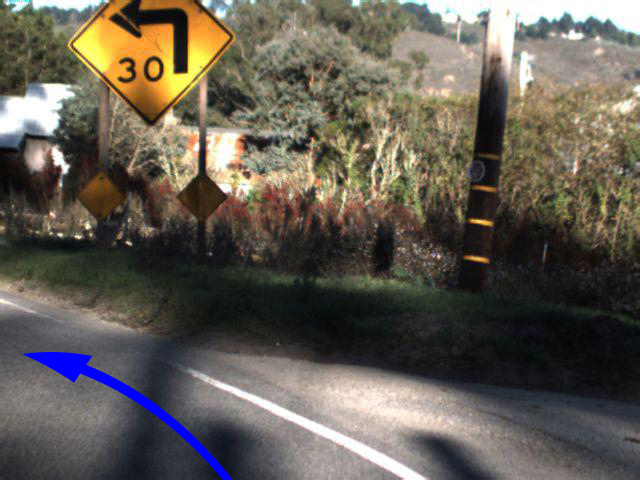}}{\scriptsize \textbf{original}}
\stackunder[4pt]{
\includegraphics[width=0.09\textwidth, height=0.09\textwidth]{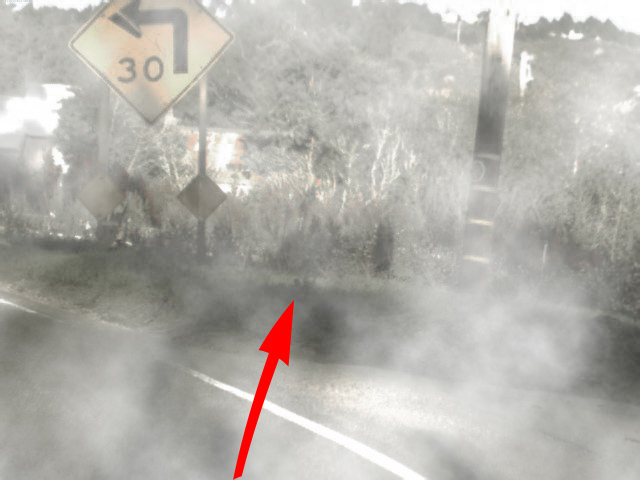}}{\scriptsize {\bf fog}}
\stackunder[4pt]{
\includegraphics[width=0.09\textwidth, height=0.09\textwidth]{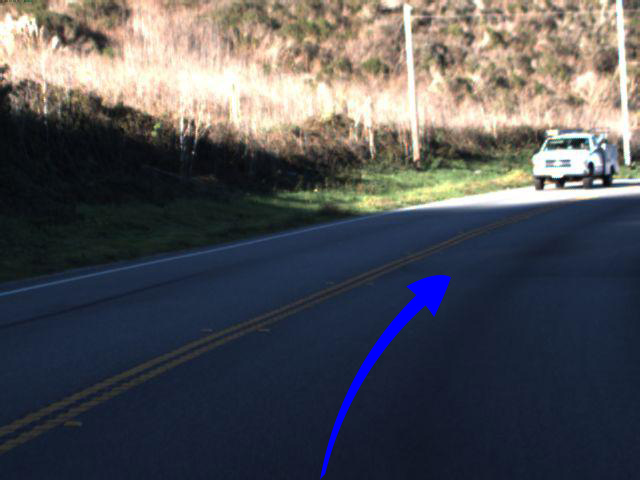}}{\scriptsize \textbf{original}}
\stackunder[4pt]{
\includegraphics[width=0.09\textwidth, height=0.09\textwidth]{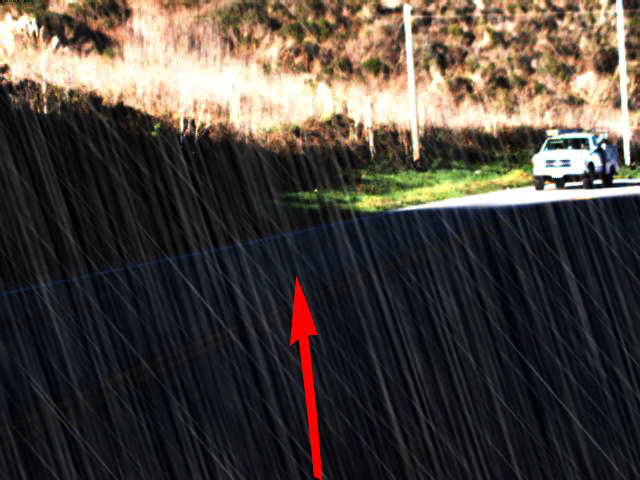}}{\scriptsize \bf rain}
\stackunder[4pt]{
\includegraphics[width=0.09\textwidth, height=0.09\textwidth]{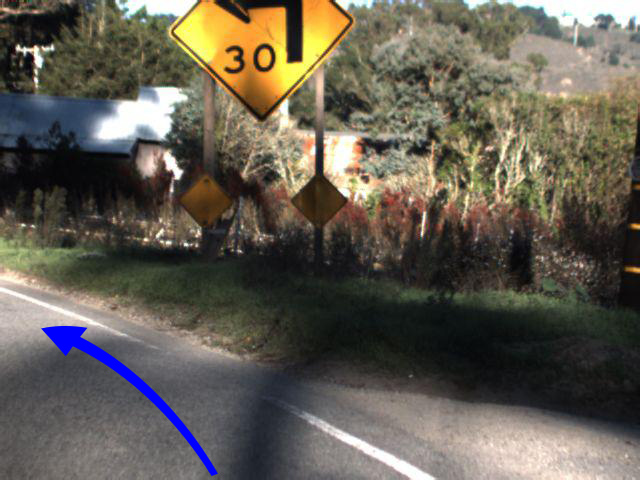}}{\scriptsize \textbf{original}}
\stackunder[4pt]{
\includegraphics[width=0.09\textwidth, height=0.09\textwidth]{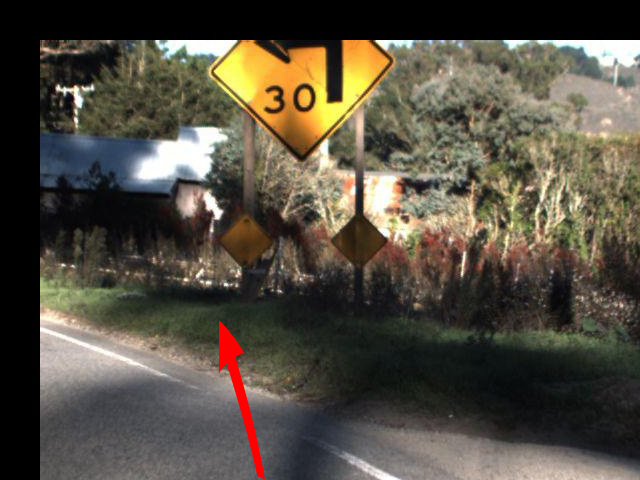}}{\scriptsize  \bf translation(40,40)}
\stackunder[4pt]{
\includegraphics[width=0.09\textwidth, height=0.09\textwidth]{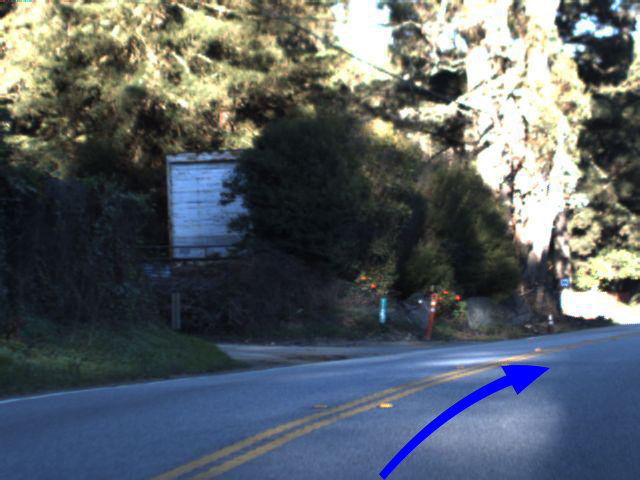}}{\scriptsize \textbf{original}}
\stackunder[4pt]{
\includegraphics[width=0.09\textwidth, height=0.09\textwidth]{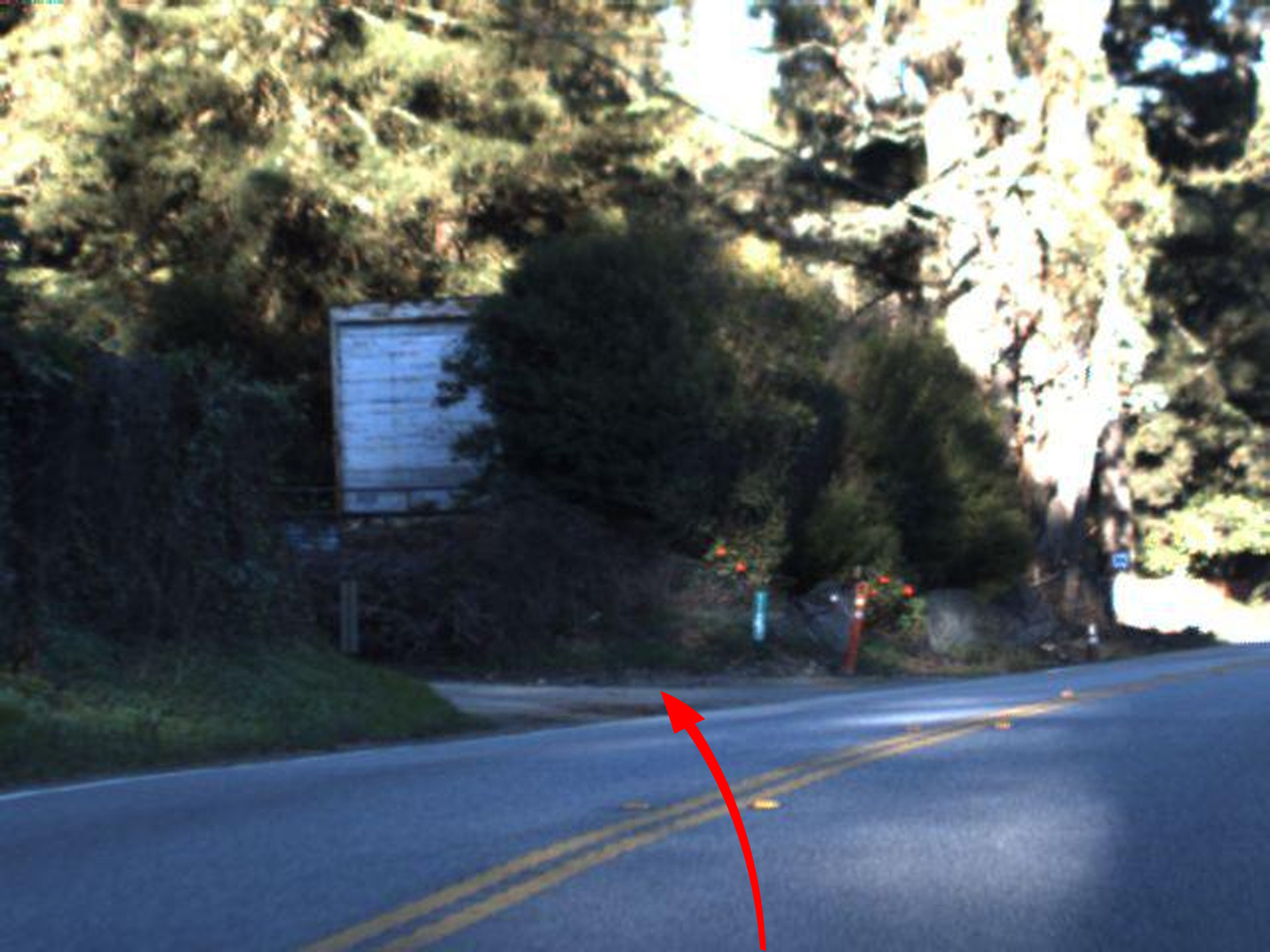}}{\scriptsize \bf scale(2.5x)}

\stackunder[4pt]{
\includegraphics[width=0.09\textwidth, height=0.09\textwidth]{{./figures/panel/1479425660724907914_arrow}.jpg}}{\scriptsize \textbf{original}}
\stackunder[4pt]{
\includegraphics[width=0.09\textwidth, height=0.09\textwidth]{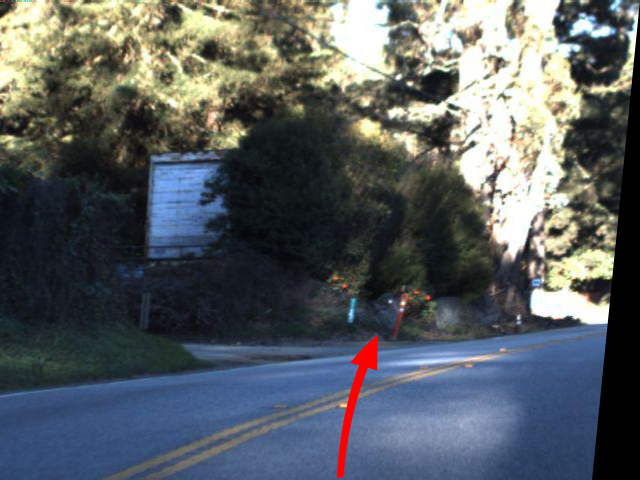}}{\scriptsize \bf shear(0.1)}
\stackunder[4pt]{
\includegraphics[width=0.09\textwidth, height=0.09\textwidth]{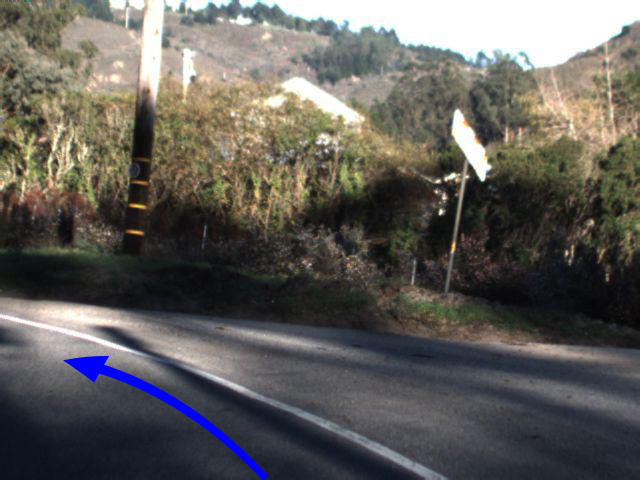}}{\scriptsize \textbf{original}}
\stackunder[4pt]{
\includegraphics[width=0.09\textwidth, height=0.09\textwidth]{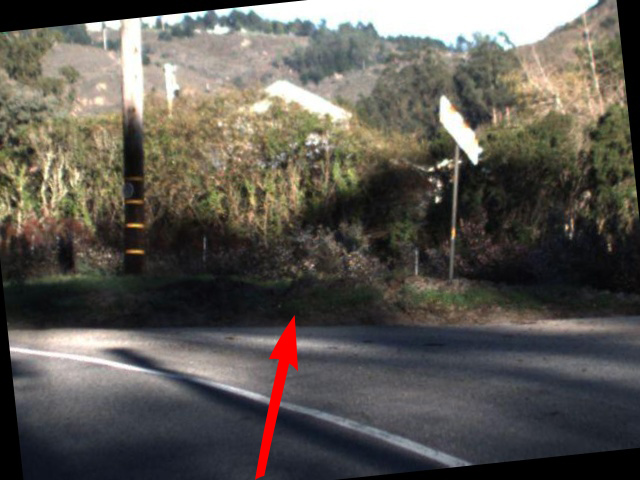}}{\scriptsize \bf rotation(6 degree)}
\stackunder[4pt]{
 \includegraphics[width=0.09\textwidth, height=0.09\textwidth]{{./figures/panel/1479425653224199015_arrow}.jpg}}{\scriptsize \textbf{original}}
\stackunder[4pt]{
\includegraphics[width=0.09\textwidth, height=0.09\textwidth]{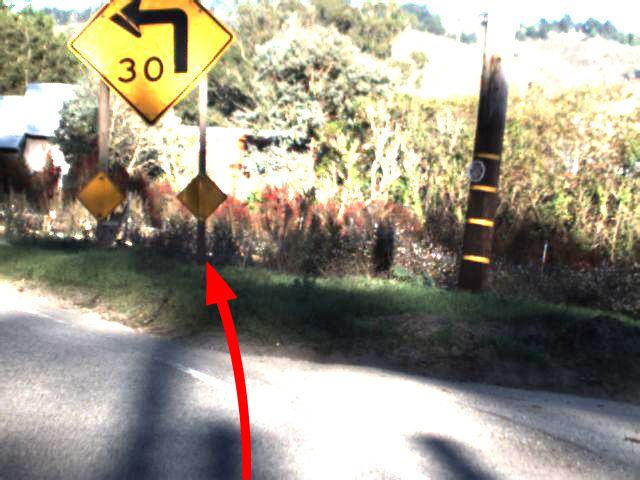}}{\scriptsize \bf contrast(1.8)}
\stackunder[4pt]{
\includegraphics[width=0.09\textwidth, height=0.09\textwidth]{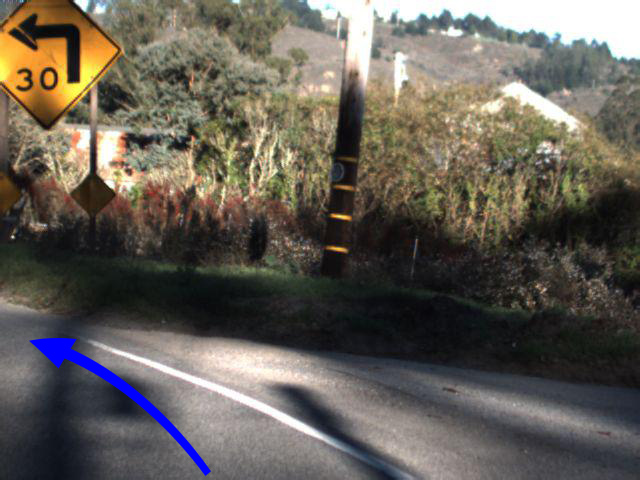}}{\scriptsize \textbf{original}}
\stackunder[4pt]{
\includegraphics[width=0.09\textwidth, height=0.09\textwidth]{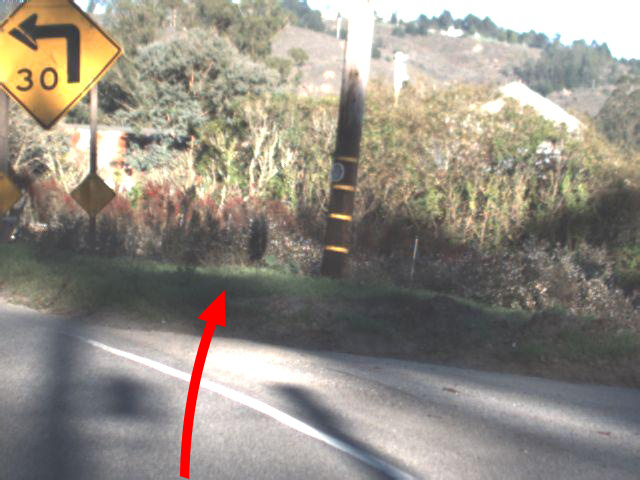}}{\scriptsize \bf brightness(50)}


\caption{\small Sample images showing erroneous behaviors detected by \pname using synthetic images.  For \textbf{original} images the arrows are marked in \blue{blue}, while for the synthetic images they are marked in \Red{red}. More such samples can be viewed at \url{https://deeplearningtest.github.io/deepTest/}.}
\label{fig:imagepanel}
\end{figure*}

However, for certain transformations (e.g., rotation), not all violations of the metamorphic relations can be considered buggy as the correct steering angle can vary widely based on the contents of the transformed image.  For example, when an image is rotated by a large amount, say 30 degrees, it is nontrivial to automatically define its correct output behavior without knowing its contents. To reduce such false positives, we only report bugs for the transformations (e.g., small rotations, rain, fog, etc.) where the correct output should not deviate much from the labels of the corresponding seed images. Thus, we further use a filtration criteria as defined in Equation~\ref{eq:filter} to identify such transformations by checking whether the MSE of the synthetic images is close to that of the original images.


\begin{equation}
\label{eq:filter}
 \mid MSE_{(trans,param)} - MSE_{org} \mid~ \le \epsilon
\end{equation}

Thus, we only choose the transformations that obey Equation~\ref{eq:filter} for counting erroneous behaviors. Table~\ref{tab:rq4Th} shows the number of such erroneous cases by varying two thresholds: $\epsilon$ and $\lambda$\textemdash a higher value of $\lambda$ and lower value of $\epsilon$ makes the system report fewer bugs and vice versa. For example, with a $\lambda$ of $5$ and $\epsilon$ of $0.03$, we report $330$ violations for simple transformations. We do not enforce the filtration criteria for composite transformations. Rain and fog effects should produce same outputs as original images. Also, in guided search since multiple transformations produce the synthesized images, it is not possible to filter out a single transformation. Thus, for rain, fog, and guided search, we report 4448, 741, and 821 erroneous behavior respectively for $\lambda=$ 5, across all three models. 


\begin{table}[!htbp]
  \centering
  \setlength{\tabcolsep}{4pt}
    \scriptsize
    \renewcommand{\arraystretch}{.8}
  \caption{\small{Number of erroneous behaviors reported by \pname across all tested models at different thresholds}}
    \label{tab:rq4Th}%
    \begin{tabular}{c|rrrrr|rrc}
    \toprule
  & \multicolumn{5}{c|}{\textbf{Simple Tranformation}} & \multicolumn{3}{l}{\textbf{Composite Transformation}} \\
     \multicolumn{1}{c|}{\bf $\lambda$}  & \multicolumn{5}{c|}{\bf $\epsilon$ (see Eqn.~\ref{eq:filter})}           & {\textbf{Fog}} & \textbf{Rain} & \multicolumn{1}{c}{\textbf{Guided}} \\
    \multicolumn{1}{c|}{\textbf{(see Eqn.~\ref{eq:mr})}} & 0.01  & 0.02  & 0.03  & 0.04  & 0.05  &       &       & \multicolumn{1}{c}{\textbf{Search}} \\
    \midrule
    1     & 15666 & 18520 & 23391 & 24952 & 29649 & 9018  & 6133  & 1148 \\
    2     & 4066  & 5033  & 6778  & 7362  & 9259  & 6503   & 2650   & 1026 \\
    3     & 1396  & 1741  & 2414  & 2627  & 3376  & 5452   & 1483   & 930 \\
    4     & 501   & 642   & 965   & 1064  & 4884  & 4884 & 997 & 872 \\
    5     & 95    & 171   & \Red{330}     & 382   & 641   & \Red{4448}    & \Red{741}     & \Red{820} \\
    6     & 49    & 85    & 185   & 210   & 359   & 4063     & 516     & 764 \\
    7     & 13    & 24    & 89    & 105   & 189   & 3732     & 287     & 721 \\
    8     & 3     & 5     & 34    & 45    & 103   & 3391     & 174     & 668 \\
    9     & 0     & 1     & 12    & 19    & 56    & 3070     & 111     & 637 \\
    10    & 0     & 0     & 3     & 5     & 23    & 2801     & 63   & 597 \\
    \bottomrule
    \end{tabular}%
\end{table}%


\begin{table}[htbp]
  \centering
  \setlength{\tabcolsep}{4pt}
    \scriptsize
    \renewcommand{\arraystretch}{.8}
	\caption{\small{Number of unique erroneous behaviors reported by \pname for different models with $\lambda=5$ (see Eqn.~\ref{eq:mr})}}
    \label{tab:rq4model}
    \begin{tabular}{l|rrr}
  \toprule
  \textbf{Transformation} & \textbf{\chauf} & \textbf{\epoch} & \textbf{\rambo} \\
  \midrule
  \underline{\bf Simple Transformation}  & & & \\
    Blur  & 3     & 27    & 11 \\
    Brightness & 97    & 32    & 15 \\
    Contrast & 31    & 12    & - \\
    Rotation & - & 13    & - \\
    Scale & - & 10    & - \\
    Shear & - & - & 23 \\
    Translation & 21    & 35    & - \\
    \midrule
    \underline{\bf Composite Transformation} & & & \\
    Rain  & 650   & 64    & 27 \\
    Fog   & 201   & 135   & 4112 \\
    Guided & 89    & 65    & 666 \\
    \bottomrule
    \end{tabular}%
    \vspace{-0.4cm}
\end{table}%


Table~\ref{tab:rq4model} further elaborates the result for different models for $\lambda=5$ and $\epsilon=0.03$, as highlighted in Table~\ref{tab:rq4Th}. Interestingly, some models are more prone to erroneous behaviors for some transformations than others. For example, \rambo produces 23 erroneous cases for shear, while the other two models do not show any such cases. Similarly, \pname finds 650 instances of erroneous behavior in \chauf for rain but only 64 and 27 for \epoch and \rambo respectively. In total, \pname detects $6339$ erroneous behaviors across all three models.
Figure~\ref{fig:imagepanel} further shows some of the erroneous behaviors that are detected by \pname under different transformations that can lead to potentially fatal situations. 

We also manually checked the bugs reported in Table~\ref{tab:rq4model} and report the false positives in Figure~\ref{fig:fp}. It also shows two synthetic images (the corresponding original images) where \pname incorrectly reports erroneous behaviors while the model's output is indeed safe. Although such manual investigation is, by definition, subjective and approximate, all the authors have reviewed the images and agreed on the false positives.



\begin{figure}[!htpb]

{\scriptsize
  \centering
    \begin{tabular}{l|ccccc}
  \toprule
     & \textbf{Simple}  &  &   &  & \\
    \textbf{Model} & \textbf{Transformation} & \textbf{Guided} &  \textbf{Rain} & \textbf{Fog} & \textbf{Total}\\
    \midrule
      \textbf{\epoch} &       14 &  0  &   0&    0 & 14\\ 
      \textbf{\chauf} &        5 &  3  &   12&   6 & 26\\
      \textbf{\rambo} &        8 &  43 &   11&   28 & 90\\ \midrule
      \textbf{Total} &        27 &  46 &   23&    34 & 130\\
 \bottomrule
    \end{tabular}%
}

\stackunder[4pt]{
\includegraphics[width=0.1\textwidth, height=0.1\textwidth]{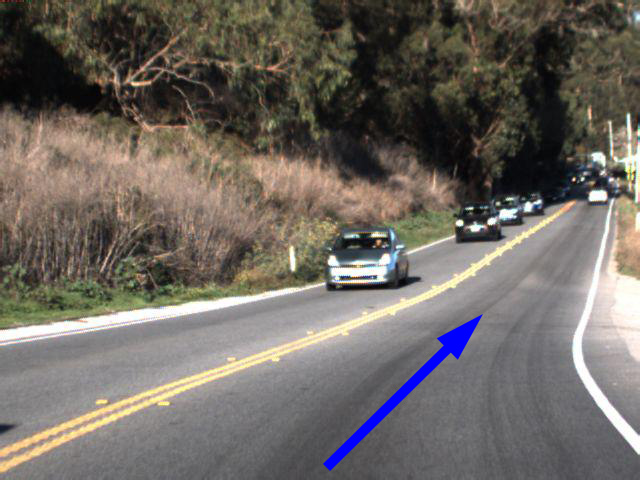}}{\scriptsize\bf original}
\stackunder[4pt]{
\includegraphics[width=0.1\textwidth, height=0.1\textwidth]{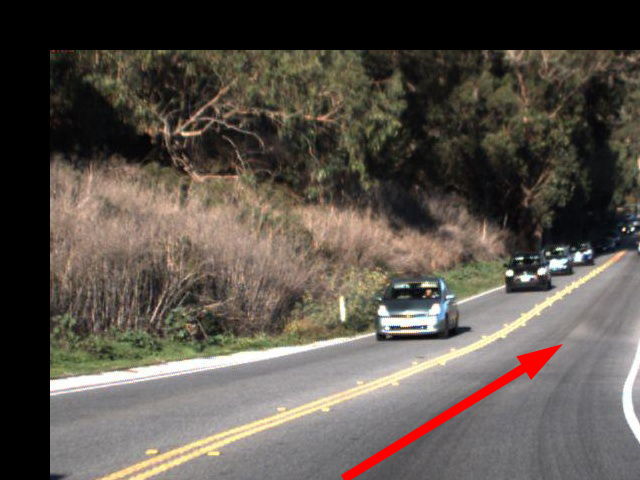}}{\scriptsize\bf translation(50,50), epoch}
\stackunder[4pt]{
\includegraphics[width=0.1\textwidth, height=0.1\textwidth]{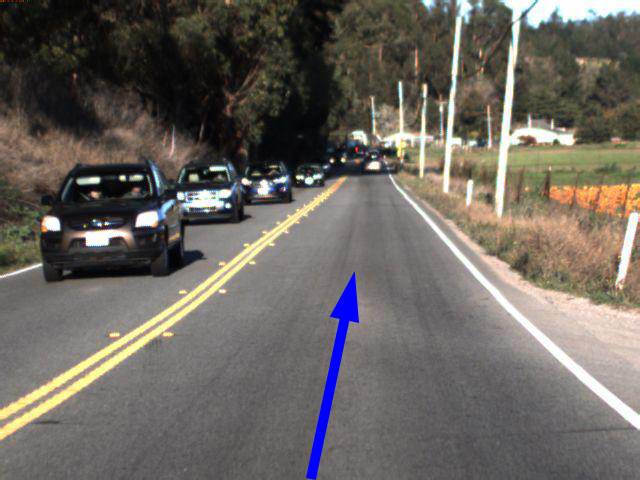}}{\scriptsize \bf original}
\stackunder[4pt]{
\includegraphics[width=0.1\textwidth, height=0.1\textwidth]{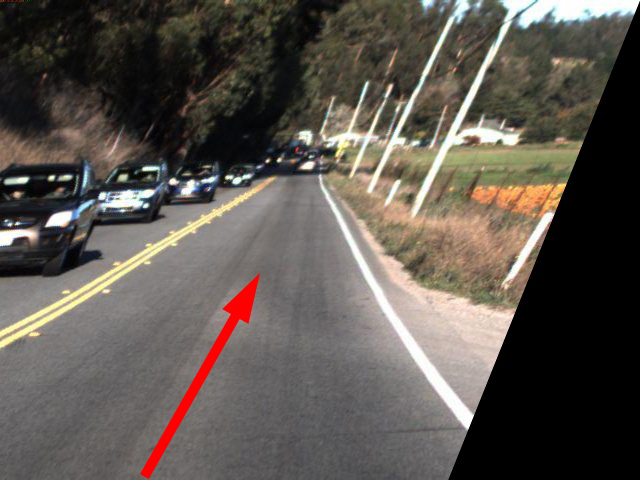}}{\scriptsize \bf shear(0.4), rambo}

\caption{\small{Sample false positives produced by \pname for $\lambda=5$, $\epsilon=0.03$}}
\label{fig:fp}
\vspace{-0.4cm}
\end{figure}

\RS{4}{With neuron coverage guided synthesized images, \pname successfully detects more than 1,000 erroneous behavior as predicted by the three models with low false positives.}   

\RQ{5}{Can retraining DNNs with synthetic images improve accuracy?}

\begin{table}[!htbp]
  \centering
  \setlength{\tabcolsep}{6pt}
    \scriptsize
    \renewcommand{\arraystretch}{.9}
  \caption{\small{Improvement in MSE after retraining of \epoch model with synthetic tests generated by \pname}}
    \begin{tabular}{lcc}
  \toprule
    Test set & \textbf{Original MSE}  &  \textbf{Retrained MSE} \\
    \midrule
      original images &       0.10 & 0.09\\ 
      with fog &        0.18 & 0.10 \\
      with rain &        0.13 &  0.07\\
 \bottomrule
    \end{tabular}%
  \label{tab:rq5}%
  \vspace{-0.2in}
\end{table}%

Here we check whether retraining the DNNs with some of the synthetic images generated by \pname helps in making the DNNs more robust.
We used the images from HMB\_3.bag~\cite{dataset} and created their synthetic versions by adding the rain and fog effects. We retrained the \epoch model with randomly sampled $66\%$ of these synthetic inputs along with the original training data. We evaluated both the original and the retrained model on 
the rest $34\%$ of the synthetic images and their original versions. In all cases, the accuracy of the retrained model improved significantly over the original model as shown in Table~\ref{tab:rq5}. 

\RS{5}{Accuracy of a DNN can be improved up to 46\% by retraining the DNN with synthetic data generated by \pname.}

\section{Threats to Validity}
\label{sec:threats}


\pname generates realistic synthetic images by applying different image transformations on the seed images. However, these transformations are not designed to be exhaustive and therefore may not cover all realistic cases.

While our transformations like rain and fog effects are designed to be realistic, the generated pictures may not be exactly reproducible in reality due to a large number of unpredictable factors, e.g., the position of the sun, the angle and size of the rain drops. etc. However, as the image processing techniques become more sophisticated, the generated pictures will get closer to reality.

A complete DNN model for driving an autonomous vehicle must also handle braking and acceleration besides the steering angle. We restricted ourselves to only test the accuracy of the steering angle as our tested models do not support braking and acceleration yet. However, our techniques should be readily applicable to testing those outputs too assuming that the models support them.

\section{Related Work}
\label{sec:rel}
{\bf Testing of driver assistance systems.} Abdessalem et al. proposed a technique for testing Advanced Driver Assistance Systems (ADAS) in autonomous cars that show warnings to the drivers if it detects pedestrians in positions with low driver visibility~\cite{abdessalem2016testing}. They use multi-objective meta heuristic search algorithms to 
generate tests that simultaneously focus on the most critical behaviors of the system and the environment as decided by the domain experts (\eg moving pedestrians in the dark). 

The key differences between this work and ours are threefold: (i) We focus on testing the image recognition and steering logic in the autonomous car DNNs while their technique tested ADAS system's warning logic based on preprocessed sensor inputs; (ii) Their blackbox technique depends on manually selected critical scenarios while our gray-box technique looks inside the DNN model and systematically maximize neuron coverage. The trade-off is that their technique can, in theory, work for arbitrary implementations while our technique is tailored for DNNs; and (iii) We leverage metamorphic relations for creating a test oracle while they depend on manual specifications for detecting faulty behavior. 

{\bf Testing and verification of machine learning.}
Traditional practices in evaluating machine learning systems primarily measure their accuracy on randomly drawn test inputs from manually labeled datasets and ad hoc simulations~\cite{witten2016data,waymo_simulation, waymoreport}. However, without the knowledge of the model's internals, such blackbox testing paradigms are not able to find different corner-cases that may induce unexpected behaviors~\cite{verify-test, pei2017deepxplore}.


Pei \etal ~\cite{pei2017deepxplore} proposed DeepXplore, a whitebox differential testing algorithm for systematically finding inputs that can trigger inconsistencies between multiple DNNs. They introduced neuron coverage as a systematic metric for measuring how much of the internal logic of a DNNs have been tested. By contrast, our graybox methods use neuron coverage for guided test generation in a single DNN and leverage metamorphic relations to identify erroneous behaviors without requiring multiple DNNs.  

Another recent line of work has explored the possibility of verifying DNNs against different safety properties ~\cite{pulina2010abstraction, huang2017safety, katz2017reluplex}. However, none of these techniques can verify a rich set of properties for real-world-sized DNNs.  By contrast, our techniques can systematically test state-of-the-art DNNs for safety-critical erroneous behaviors but do not provide any theoretical guarantees.


{\bf Adversarial machine learning.}
A large number of projects successfully attacked machine learning models at test time by forcing it to make unexpected mistakes. More specifically, these attacks focus on finding inputs that, when changed minimally from their original versions, get classified differently by the machine learning classifiers. These types of attacks are known to affect a broad spectrum of tasks such as image recognition~\cite{szegedy2013intriguing, goodfellow2014explaining, narodytska2016simple, nguyen2015deep, liu2016delving, papernot2016limitations, papernot2017practical, kos2017adversarial, evtimov2017robust}, face detection/verification~\cite{wilber2016can, sharif2016accessorize}, malware detection~\cite{laskov2014practical, xu2016automatically, grosse2016adversarial, endgame-evade}, and text analysis~\cite{papernot2016crafting, miyato2016adversarial}. Several prior works have explored defenses against these attacks with different effectiveness~\cite{gu2014towards, shaham2015understanding, bastani2016measuring, zheng2016improving, huang2017safety, carlini2017towards, cisse2017parseval, wang2014man, papernot2016distillation, xu2017feature, grosse2017statistical, steinhardt2017certified, feinman2017detecting, metzen2017detecting, papernot2017extending}.




In summary, this line of work tries to find a particular class of erroneous behaviors, i.e., forcing incorrect prediction by adding a minimum amount of noise to a given input. By contrast, we systematically test a given DNN by maximizing neuron coverage and find a diverse set of corner-case behaviors. Moreover, we specifically focus on finding realistic conditions that can occur in practice.   


{\bf Test amplification.} 
There is a large body of work on test case generation and amplification techniques for traditional software that automatically generate test cases from some seed inputs and increase code coverage. Instead of summarizing them individually here, we refer the interested readers to the surveys by Anand et al.~\cite{anand2013orchestrated}, McMinn et al.~\cite{mcminn2004search}, and Pasareanu et al.~\cite{puasuareanu2009survey}. Unlike these approaches, \pname is designed to operate on DNNs. 

{\bf Metamorphic testing.} Metamorphic testing \cite{chen1998metamorphic, zhou2004metamorphic} is a way of creating test oracles in settings where manual specifications are not available. Metamorphic testing identifies buggy behavior by detecting violations of domain-specific metamorphic relations that are defined across outputs from multiple executions of the test program with different inputs.  For example, a sample metamorphic property for program $p$ adding two inputs $a$ and $b$ can be $p(a,b)=p(a,0)+p(b,0)$. Metamorphic testing has been used in the past for testing both supervised and unsupervised machine learning classifiers~\cite{murphy2008properties, xie2009application}. By contrast, we define new metamorphic relations in the domain of autonomous cars which, unlike the classifiers tested before, produce a continuous steering angle, i.e., it is a regression task. 

\section{Conclusion}
\label{sec:conclusion}

In this paper, we proposed and evaluated \pname, a tool for automated testing of DNN-driven autonomous cars. 
\pname maximizes the neuron coverage of a DNN using synthetic test images generated by applying different realistic transformations on a set of seed images. We use domain-specific metamorphic relations to find erroneous behaviors of the DNN without detailed specification. \pname can be easily adapted to test other DNN-based systems by customizing the transformations and metamorphic relations. We believe \pname is an important first step towards building robust DNN-based systems.

\section{Acknowledgements}
We would like to thank Yoav Hollander and the anonymous reviewers for their helpful feedback. This work was supported in part by NSF grants CNS-16-18771, CNS-16-17670, and CNS-15-64055; ONR grant N00014-17-1-2010; and
a Google Faculty Fellowship.


\balance

\bibliographystyle{ACM-Reference-Format}
\bibliography{main} 


\begin{thebibliography}{87}


\ifx \showCODEN    \undefined \def \showCODEN     #1{\unskip}     \fi
\ifx \showDOI      \undefined \def \showDOI       #1{#1}\fi
\ifx \showISBNx    \undefined \def \showISBNx     #1{\unskip}     \fi
\ifx \showISBNxiii \undefined \def \showISBNxiii  #1{\unskip}     \fi
\ifx \showISSN     \undefined \def \showISSN      #1{\unskip}     \fi
\ifx \showLCCN     \undefined \def \showLCCN      #1{\unskip}     \fi
\ifx \shownote     \undefined \def \shownote      #1{#1}          \fi
\ifx \showarticletitle \undefined \def \showarticletitle #1{#1}   \fi
\ifx \showURL      \undefined \def \showURL       {\relax}        \fi
\providecommand\bibfield[2]{#2}
\providecommand\bibinfo[2]{#2}
\providecommand\natexlab[1]{#1}
\providecommand\showeprint[2][]{arXiv:#2}

\bibitem[\protect\citeauthoryear{??}{rai}{2013}]%
        {raineffect}
 \bibinfo{year}{2013}\natexlab{}.
\newblock \bibinfo{title}{Add Dramatic Rain to a Photo in Photoshop}.
\newblock
  \bibinfo{howpublished}{\url{https://design.tutsplus.com/tutorials/add-dramatic-rain-to-a-photo-in-photoshop--psd-29536}}.
    (\bibinfo{year}{2013}).
\newblock


\bibitem[\protect\citeauthoryear{??}{fog}{2013}]%
        {fogeffect}
 \bibinfo{year}{2013}\natexlab{}.
\newblock \bibinfo{title}{How to create mist: Photoshop effects for atmospheric
  landscapes}.
\newblock
  \bibinfo{howpublished}{\url{http://www.techradar.com/how-to/photography-video-capture/cameras/how-to-create-mist-photoshop-effects-for-atmospheric-landscapes-1320997}}.
    (\bibinfo{year}{2013}).
\newblock


\bibitem[\protect\citeauthoryear{??}{its}{2014}]%
        {itseez2014theopencv}
 \bibinfo{year}{2014}\natexlab{}.
\newblock \bibinfo{booktitle}{\emph{The OpenCV Reference Manual}
  (\bibinfo{edition}{2.4.9.0} ed.)}.
\newblock


\bibitem[\protect\citeauthoryear{??}{hyu}{2014}]%
        {hyundai.crash}
 \bibinfo{year}{2014}\natexlab{}.
\newblock \bibinfo{title}{This Is How Bad Self-Driving Cars Suck In The Rain}.
\newblock
  \bibinfo{howpublished}{\url{http://jalopnik.com/this-is-how-bad-self-driving-cars-suck-in-the-rain-1666268433}}.
    (\bibinfo{year}{2014}).
\newblock


\bibitem[\protect\citeauthoryear{??}{aff}{2015a}]%
        {affinetransform1}
 \bibinfo{year}{2015}\natexlab{a}.
\newblock \bibinfo{title}{Affine Transformation}.
\newblock
  \bibinfo{howpublished}{\url{https://www.mathworks.com/discovery/affine-transformation.html}}.
    (\bibinfo{year}{2015}).
\newblock


\bibitem[\protect\citeauthoryear{??}{aff}{2015b}]%
        {affinetransform2}
 \bibinfo{year}{2015}\natexlab{b}.
\newblock \bibinfo{title}{Affine Transformations}.
\newblock
  \bibinfo{howpublished}{\url{http://docs.opencv.org/3.1.0/d4/d61/tutorial_warp_affine.html}}.
    (\bibinfo{year}{2015}).
\newblock


\bibitem[\protect\citeauthoryear{??}{its}{2015}]%
        {itseez2015opencv}
 \bibinfo{year}{2015}\natexlab{}.
\newblock \bibinfo{title}{Open Source Computer Vision Library}.
\newblock \bibinfo{howpublished}{\url{https://github.com/itseez/opencv}}.
  (\bibinfo{year}{2015}).
\newblock


\bibitem[\protect\citeauthoryear{??}{cha}{2016a}]%
        {chauffeur}
 \bibinfo{year}{2016}\natexlab{a}.
\newblock \bibinfo{title}{Chauffeur model}.
\newblock
  \bibinfo{howpublished}{\url{https://github.com/udacity/self-driving-car/tree/master/steering-models/community-models/chauffeur}}.
    (\bibinfo{year}{2016}).
\newblock


\bibitem[\protect\citeauthoryear{??}{com}{2016}]%
        {comma.ai}
 \bibinfo{year}{2016}\natexlab{}.
\newblock \bibinfo{title}{comma.ai's steering model}.
\newblock
  \bibinfo{howpublished}{\url{https://github.com/commaai/research/blob/master/train_steering_model.py}}.
    (\bibinfo{year}{2016}).
\newblock


\bibitem[\protect\citeauthoryear{??}{epo}{2016}]%
        {epoch}
 \bibinfo{year}{2016}\natexlab{}.
\newblock \bibinfo{title}{Epoch model}.
\newblock
  \bibinfo{howpublished}{\url{https://github.com/udacity/self-driving-car/tree/master/steering-models/community-models/cg23}}.
    (\bibinfo{year}{2016}).
\newblock


\bibitem[\protect\citeauthoryear{??}{way}{2016}]%
        {waymoreport}
 \bibinfo{year}{2016}\natexlab{}.
\newblock \bibinfo{title}{Google Auto Waymo Disengagement Report for Autonomous
  Driving}.
\newblock
  \bibinfo{howpublished}{\url{https://www.dmv.ca.gov/portal/wcm/connect/946b3502-c959-4e3b-b119-91319c27788f/GoogleAutoWaymo_disengage_report_2016.pdf?MOD=AJPERES}}.
    (\bibinfo{year}{2016}).
\newblock


\bibitem[\protect\citeauthoryear{??}{goo}{2016}]%
        {google.crash}
 \bibinfo{year}{2016}\natexlab{}.
\newblock \bibinfo{title}{Google's Self-Driving Car Caused Its First Crash}.
\newblock
  \bibinfo{howpublished}{\url{https://www.wired.com/2016/02/googles-self-driving-car-may-caused-first-crash/}}.
    (\bibinfo{year}{2016}).
\newblock


\bibitem[\protect\citeauthoryear{??}{ram}{2016}]%
        {rambo}
 \bibinfo{year}{2016}\natexlab{}.
\newblock \bibinfo{title}{Rambo model}.
\newblock
  \bibinfo{howpublished}{\url{https://github.com/udacity/self-driving-car/tree/master/steering-models/community-models/rambo}}.
    (\bibinfo{year}{2016}).
\newblock


\bibitem[\protect\citeauthoryear{??}{aut}{2016}]%
        {autopilot}
 \bibinfo{year}{2016}\natexlab{}.
\newblock \bibinfo{title}{Tesla Autopilot}.
\newblock \bibinfo{howpublished}{\url{https://www.tesla.com/autopilot}}.
  (\bibinfo{year}{2016}).
\newblock


\bibitem[\protect\citeauthoryear{??}{cha}{2016b}]%
        {challenge2}
 \bibinfo{year}{2016}\natexlab{b}.
\newblock \bibinfo{title}{Udacity self driving car challenge 2}.
\newblock
  \bibinfo{howpublished}{\url{https://github.com/udacity/self-driving-car/tree/master/challenges/challenge-2}}.
    (\bibinfo{year}{2016}).
\newblock


\bibitem[\protect\citeauthoryear{??}{dat}{2016}]%
        {dataset}
 \bibinfo{year}{2016}\natexlab{}.
\newblock \bibinfo{title}{Udacity self driving car challenge 2 dataset}.
\newblock
  \bibinfo{howpublished}{\url{https://github.com/udacity/self-driving-car/tree/master/datasets/CH2}}.
    (\bibinfo{year}{2016}).
\newblock


\bibitem[\protect\citeauthoryear{??}{cnn}{2016}]%
        {cnn.crash}
 \bibinfo{year}{2016}\natexlab{}.
\newblock \bibinfo{title}{Who's responsible when an autonomous car crashes?}
\newblock
  \bibinfo{howpublished}{\url{http://money.cnn.com/2016/07/07/technology/tesla-liability-risk/index.html}}.
    (\bibinfo{year}{2016}).
\newblock


\bibitem[\protect\citeauthoryear{??}{sel}{2017a}]%
        {selfdrivingcar4}
 \bibinfo{year}{2017}\natexlab{a}.
\newblock \bibinfo{title}{Autonomous Vehicles Enacted Legislation}.
\newblock
  \bibinfo{howpublished}{\url{http://www.ncsl.org/research/transportation/autonomous-vehicles-self-driving-vehicles-enacted-legislation.aspx}}.
    (\bibinfo{year}{2017}).
\newblock


\bibitem[\protect\citeauthoryear{??}{apo}{2017}]%
        {apollo}
 \bibinfo{year}{2017}\natexlab{}.
\newblock \bibinfo{title}{Baidu Apollo}.
\newblock \bibinfo{howpublished}{\url{https://github.com/ApolloAuto/apollo}}.
  (\bibinfo{year}{2017}).
\newblock


\bibitem[\protect\citeauthoryear{??}{way}{2017}]%
        {waymo_simulation}
 \bibinfo{year}{2017}\natexlab{}.
\newblock \bibinfo{title}{Inside Waymo's Secret World for Training Self-Driving
  Cars}.
\newblock
  \bibinfo{howpublished}{\url{https://www.theatlantic.com/technology/archive/2017/08/inside-waymos-secret-testing-and-simulation-facilities/537648/}}.
    (\bibinfo{year}{2017}).
\newblock


\bibitem[\protect\citeauthoryear{??}{sel}{2017b}]%
        {selfdrivingcar1}
 \bibinfo{year}{2017}\natexlab{b}.
\newblock \bibinfo{title}{The Numbers Don't Lie: Self-Driving Cars Are Getting
  Good}.
\newblock
  \bibinfo{howpublished}{\url{https://www.wired.com/2017/02/california-dmv-autonomous-car-disengagement}}.
    (\bibinfo{year}{2017}).
\newblock


\bibitem[\protect\citeauthoryear{??}{sof}{2017}]%
        {software2}
 \bibinfo{year}{2017}\natexlab{}.
\newblock \bibinfo{title}{Software 2.0}.
\newblock
  \bibinfo{howpublished}{\url{https://medium.com/@karpathy/software-2-0-a64152b37c35}}.
    (\bibinfo{year}{2017}).
\newblock


\bibitem[\protect\citeauthoryear{??}{nyt}{2017}]%
        {nyt.crash}
 \bibinfo{year}{2017}\natexlab{}.
\newblock \bibinfo{title}{Tesla's Self-Driving System Cleared in Deadly Crash}.
\newblock
  \bibinfo{howpublished}{\url{https://www.nytimes.com/2017/01/19/business/tesla-model-s-autopilot-fatal-crash.html}}.
    (\bibinfo{year}{2017}).
\newblock


\bibitem[\protect\citeauthoryear{Abadi, Agarwal, Barham, Brevdo, Chen, Citro,
  Corrado, Davis, Dean, Devin, et~al\mbox{.}}{Abadi et~al\mbox{.}}{2016}]%
        {abadi2016tensorflow}
\bibfield{author}{\bibinfo{person}{Mart{\'\i}n Abadi}, \bibinfo{person}{Ashish
  Agarwal}, \bibinfo{person}{Paul Barham}, \bibinfo{person}{Eugene Brevdo},
  \bibinfo{person}{Zhifeng Chen}, \bibinfo{person}{Craig Citro},
  \bibinfo{person}{Greg~S Corrado}, \bibinfo{person}{Andy Davis},
  \bibinfo{person}{Jeffrey Dean}, \bibinfo{person}{Matthieu Devin},
  {et~al\mbox{.}}} \bibinfo{year}{2016}\natexlab{}.
\newblock \showarticletitle{Tensorflow: Large-scale machine learning on
  heterogeneous distributed systems}.
\newblock \bibinfo{journal}{\emph{arXiv preprint arXiv:1603.04467}}
  (\bibinfo{year}{2016}).
\newblock


\bibitem[\protect\citeauthoryear{Abdessalem, Nejati, Briand, and
  Stifter}{Abdessalem et~al\mbox{.}}{2016}]%
        {abdessalem2016testing}
\bibfield{author}{\bibinfo{person}{Raja~Ben Abdessalem}, \bibinfo{person}{Shiva
  Nejati}, \bibinfo{person}{Lionel~C Briand}, {and} \bibinfo{person}{Thomas
  Stifter}.} \bibinfo{year}{2016}\natexlab{}.
\newblock \showarticletitle{Testing advanced driver assistance systems using
  multi-objective search and neural networks}. In
  \bibinfo{booktitle}{\emph{Automated Software Engineering (ASE), 2016 31st
  IEEE/ACM International Conference on}}. IEEE, \bibinfo{pages}{63--74}.
\newblock


\bibitem[\protect\citeauthoryear{an~Goodfellow and Papernot}{an~Goodfellow and
  Papernot}{2017}]%
        {verify-test}
\bibfield{author}{\bibinfo{person}{an Goodfellow} {and}
  \bibinfo{person}{Nicolas Papernot}.} \bibinfo{year}{2017}\natexlab{}.
\newblock \bibinfo{title}{The challenge of verification and testing of machine
  learning}.
\newblock
  \bibinfo{howpublished}{\url{http://www.cleverhans.io/security/privacy/ml/2017/06/14/verification.html}}.
    (\bibinfo{year}{2017}).
\newblock


\bibitem[\protect\citeauthoryear{Anand, Burke, Chen, Clark, Cohen, Grieskamp,
  Harman, Harrold, Mcminn, Bertolino, et~al\mbox{.}}{Anand
  et~al\mbox{.}}{2013}]%
        {anand2013orchestrated}
\bibfield{author}{\bibinfo{person}{Saswat Anand}, \bibinfo{person}{Edmund~K
  Burke}, \bibinfo{person}{Tsong~Yueh Chen}, \bibinfo{person}{John Clark},
  \bibinfo{person}{Myra~B Cohen}, \bibinfo{person}{Wolfgang Grieskamp},
  \bibinfo{person}{Mark Harman}, \bibinfo{person}{Mary~Jean Harrold},
  \bibinfo{person}{Phil Mcminn}, \bibinfo{person}{Antonia Bertolino},
  {et~al\mbox{.}}} \bibinfo{year}{2013}\natexlab{}.
\newblock \showarticletitle{An orchestrated survey of methodologies for
  automated software test case generation}.
\newblock \bibinfo{journal}{\emph{Journal of Systems and Software}}
  \bibinfo{volume}{86}, \bibinfo{number}{8} (\bibinfo{year}{2013}),
  \bibinfo{pages}{1978--2001}.
\newblock


\bibitem[\protect\citeauthoryear{Anderson}{Anderson}{2017}]%
        {endgame-evade}
\bibfield{author}{\bibinfo{person}{Hyrum Anderson}.}
  \bibinfo{year}{2017}\natexlab{}.
\newblock \bibinfo{title}{Evading Next-Gen AV using A.I.}
\newblock
  \bibinfo{howpublished}{\url{https://www.defcon.org/html/defcon-25/dc-25-index.html}}.
    (\bibinfo{year}{2017}).
\newblock


\bibitem[\protect\citeauthoryear{Bastani, Ioannou, Lampropoulos, Vytiniotis,
  Nori, and Criminisi}{Bastani et~al\mbox{.}}{2016}]%
        {bastani2016measuring}
\bibfield{author}{\bibinfo{person}{Osbert Bastani}, \bibinfo{person}{Yani
  Ioannou}, \bibinfo{person}{Leonidas Lampropoulos}, \bibinfo{person}{Dimitrios
  Vytiniotis}, \bibinfo{person}{Aditya Nori}, {and} \bibinfo{person}{Antonio
  Criminisi}.} \bibinfo{year}{2016}\natexlab{}.
\newblock \showarticletitle{Measuring neural net robustness with constraints}.
  In \bibinfo{booktitle}{\emph{Advances in Neural Information Processing
  Systems}}. \bibinfo{pages}{2613--2621}.
\newblock


\bibitem[\protect\citeauthoryear{Bengio, Lamblin, Popovici, and
  Larochelle}{Bengio et~al\mbox{.}}{2007}]%
        {bengio2007greedy}
\bibfield{author}{\bibinfo{person}{Yoshua Bengio}, \bibinfo{person}{Pascal
  Lamblin}, \bibinfo{person}{Dan Popovici}, {and} \bibinfo{person}{Hugo
  Larochelle}.} \bibinfo{year}{2007}\natexlab{}.
\newblock \showarticletitle{Greedy layer-wise training of deep networks}. In
  \bibinfo{booktitle}{\emph{Advances in neural information processing
  systems}}. \bibinfo{pages}{153--160}.
\newblock


\bibitem[\protect\citeauthoryear{Bojarski, Del~Testa, Dworakowski, Firner,
  Flepp, Goyal, Jackel, Monfort, Muller, Zhang, et~al\mbox{.}}{Bojarski
  et~al\mbox{.}}{2016}]%
        {bojarski2016end}
\bibfield{author}{\bibinfo{person}{Mariusz Bojarski}, \bibinfo{person}{Davide
  Del~Testa}, \bibinfo{person}{Daniel Dworakowski}, \bibinfo{person}{Bernhard
  Firner}, \bibinfo{person}{Beat Flepp}, \bibinfo{person}{Prasoon Goyal},
  \bibinfo{person}{Lawrence~D Jackel}, \bibinfo{person}{Mathew Monfort},
  \bibinfo{person}{Urs Muller}, \bibinfo{person}{Jiakai Zhang},
  {et~al\mbox{.}}} \bibinfo{year}{2016}\natexlab{}.
\newblock \showarticletitle{End to end learning for self-driving cars}.
\newblock \bibinfo{journal}{\emph{arXiv preprint arXiv:1604.07316}}
  (\bibinfo{year}{2016}).
\newblock


\bibitem[\protect\citeauthoryear{Carlini and Wagner}{Carlini and
  Wagner}{2017}]%
        {carlini2017towards}
\bibfield{author}{\bibinfo{person}{Nicholas Carlini} {and}
  \bibinfo{person}{David Wagner}.} \bibinfo{year}{2017}\natexlab{}.
\newblock \showarticletitle{Towards evaluating the robustness of neural
  networks}. In \bibinfo{booktitle}{\emph{Security and Privacy (SP), 2017 IEEE
  Symposium on}}. IEEE, \bibinfo{pages}{39--57}.
\newblock


\bibitem[\protect\citeauthoryear{Chen, Cheung, and Yiu}{Chen
  et~al\mbox{.}}{1998}]%
        {chen1998metamorphic}
\bibfield{author}{\bibinfo{person}{Tsong~Y Chen}, \bibinfo{person}{Shing~C
  Cheung}, {and} \bibinfo{person}{Shiu~Ming Yiu}.}
  \bibinfo{year}{1998}\natexlab{}.
\newblock \bibinfo{booktitle}{\emph{Metamorphic testing: a new approach for
  generating next test cases}}.
\newblock \bibinfo{type}{{T}echnical {R}eport}. \bibinfo{institution}{Technical
  Report HKUST-CS98-01, Department of Computer Science, Hong Kong University of
  Science and Technology, Hong Kong}.
\newblock


\bibitem[\protect\citeauthoryear{Chollet et~al\mbox{.}}{Chollet
  et~al\mbox{.}}{2015}]%
        {chollet2015keras}
\bibfield{author}{\bibinfo{person}{Fran\c{c}ois Chollet} {et~al\mbox{.}}}
  \bibinfo{year}{2015}\natexlab{}.
\newblock \bibinfo{title}{Keras}.
\newblock \bibinfo{howpublished}{\url{https://github.com/fchollet/keras}}.
  (\bibinfo{year}{2015}).
\newblock


\bibitem[\protect\citeauthoryear{Cisse, Bojanowski, Grave, Dauphin, and
  Usunier}{Cisse et~al\mbox{.}}{2017}]%
        {cisse2017parseval}
\bibfield{author}{\bibinfo{person}{Moustapha Cisse}, \bibinfo{person}{Piotr
  Bojanowski}, \bibinfo{person}{Edouard Grave}, \bibinfo{person}{Yann Dauphin},
  {and} \bibinfo{person}{Nicolas Usunier}.} \bibinfo{year}{2017}\natexlab{}.
\newblock \showarticletitle{Parseval networks: Improving robustness to
  adversarial examples}. In \bibinfo{booktitle}{\emph{International Conference
  on Machine Learning}}. \bibinfo{pages}{854--863}.
\newblock


\bibitem[\protect\citeauthoryear{DMV}{DMV}{2016}]%
        {selfdrivingcar2}
\bibfield{author}{\bibinfo{person}{California DMV}.}
  \bibinfo{year}{2016}\natexlab{}.
\newblock \bibinfo{title}{Autonomous Vehicle Disengagement Reports}.
\newblock
  \bibinfo{howpublished}{\url{https://www.dmv.ca.gov/portal/dmv/detail/vr/autonomous/disengagement_report_2016}}.
    (\bibinfo{year}{2016}).
\newblock


\bibitem[\protect\citeauthoryear{Evtimov, Eykholt, Fernandes, Kohno, Li,
  Prakash, Rahmati, and Song}{Evtimov et~al\mbox{.}}{2017}]%
        {evtimov2017robust}
\bibfield{author}{\bibinfo{person}{Ivan Evtimov}, \bibinfo{person}{Kevin
  Eykholt}, \bibinfo{person}{Earlence Fernandes}, \bibinfo{person}{Tadayoshi
  Kohno}, \bibinfo{person}{Bo Li}, \bibinfo{person}{Atul Prakash},
  \bibinfo{person}{Amir Rahmati}, {and} \bibinfo{person}{Dawn Song}.}
  \bibinfo{year}{2017}\natexlab{}.
\newblock \showarticletitle{Robust Physical-World Attacks on Machine Learning
  Models}.
\newblock \bibinfo{journal}{\emph{arXiv preprint arXiv:1707.08945}}
  (\bibinfo{year}{2017}).
\newblock


\bibitem[\protect\citeauthoryear{Feinman, Curtin, Shintre, and Gardner}{Feinman
  et~al\mbox{.}}{2017}]%
        {feinman2017detecting}
\bibfield{author}{\bibinfo{person}{Reuben Feinman}, \bibinfo{person}{Ryan~R
  Curtin}, \bibinfo{person}{Saurabh Shintre}, {and} \bibinfo{person}{Andrew~B
  Gardner}.} \bibinfo{year}{2017}\natexlab{}.
\newblock \showarticletitle{Detecting Adversarial Samples from Artifacts}.
\newblock \bibinfo{journal}{\emph{arXiv preprint arXiv:1703.00410}}
  (\bibinfo{year}{2017}).
\newblock


\bibitem[\protect\citeauthoryear{Goodfellow, Bengio, and Courville}{Goodfellow
  et~al\mbox{.}}{2016}]%
        {Goodfellow-et-al-2016-Book}
\bibfield{author}{\bibinfo{person}{Ian Goodfellow}, \bibinfo{person}{Yoshua
  Bengio}, {and} \bibinfo{person}{Aaron Courville}.}
  \bibinfo{year}{2016}\natexlab{}.
\newblock \bibinfo{booktitle}{\emph{Deep Learning}}.
\newblock
\urldef\tempurl%
\url{http://www.deeplearningbook.org}
\showURL{%
\tempurl}
\newblock
\shownote{Book in preparation for MIT Press.}


\bibitem[\protect\citeauthoryear{Goodfellow, Shlens, and Szegedy}{Goodfellow
  et~al\mbox{.}}{2015}]%
        {goodfellow2014explaining}
\bibfield{author}{\bibinfo{person}{Ian~J Goodfellow}, \bibinfo{person}{Jonathon
  Shlens}, {and} \bibinfo{person}{Christian Szegedy}.}
  \bibinfo{year}{2015}\natexlab{}.
\newblock \showarticletitle{Explaining and harnessing adversarial examples}. In
  \bibinfo{booktitle}{\emph{International Conference on Learning
  Representations (ICLR)}}.
\newblock


\bibitem[\protect\citeauthoryear{Grosse, Manoharan, Papernot, Backes, and
  McDaniel}{Grosse et~al\mbox{.}}{2017a}]%
        {grosse2017statistical}
\bibfield{author}{\bibinfo{person}{Kathrin Grosse}, \bibinfo{person}{Praveen
  Manoharan}, \bibinfo{person}{Nicolas Papernot}, \bibinfo{person}{Michael
  Backes}, {and} \bibinfo{person}{Patrick McDaniel}.}
  \bibinfo{year}{2017}\natexlab{a}.
\newblock \showarticletitle{On the (statistical) detection of adversarial
  examples}.
\newblock \bibinfo{journal}{\emph{arXiv preprint arXiv:1702.06280}}
  (\bibinfo{year}{2017}).
\newblock


\bibitem[\protect\citeauthoryear{Grosse, Papernot, Manoharan, Backes, and
  McDaniel}{Grosse et~al\mbox{.}}{2017b}]%
        {grosse2016adversarial}
\bibfield{author}{\bibinfo{person}{Kathrin Grosse}, \bibinfo{person}{Nicolas
  Papernot}, \bibinfo{person}{Praveen Manoharan}, \bibinfo{person}{Michael
  Backes}, {and} \bibinfo{person}{Patrick~D. McDaniel}.}
  \bibinfo{year}{2017}\natexlab{b}.
\newblock \showarticletitle{Adversarial Perturbations Against Deep Neural
  Networks for Malware Classification}. In
  \bibinfo{booktitle}{\emph{Proceedings of the 2017 European Symposium on
  Research in Computer Security}}.
\newblock


\bibitem[\protect\citeauthoryear{Gu and Rigazio}{Gu and Rigazio}{2015}]%
        {gu2014towards}
\bibfield{author}{\bibinfo{person}{Shixiang Gu} {and} \bibinfo{person}{Luca
  Rigazio}.} \bibinfo{year}{2015}\natexlab{}.
\newblock \showarticletitle{Towards deep neural network architectures robust to
  adversarial examples}. In \bibinfo{booktitle}{\emph{International Conference
  on Learning Representations (ICLR)}}.
\newblock


\bibitem[\protect\citeauthoryear{Hauke and Kossowski}{Hauke and
  Kossowski}{2011}]%
        {hauke2011comparison}
\bibfield{author}{\bibinfo{person}{Jan Hauke} {and} \bibinfo{person}{Tomasz
  Kossowski}.} \bibinfo{year}{2011}\natexlab{}.
\newblock \showarticletitle{Comparison of values of Pearson's and Spearman's
  correlation coefficients on the same sets of data}.
\newblock \bibinfo{journal}{\emph{Quaestiones geographicae}}
  \bibinfo{volume}{30}, \bibinfo{number}{2} (\bibinfo{year}{2011}),
  \bibinfo{pages}{87}.
\newblock


\bibitem[\protect\citeauthoryear{Hijazi, Kumar, and Rowen}{Hijazi
  et~al\mbox{.}}{2015}]%
        {hijazi2015using}
\bibfield{author}{\bibinfo{person}{Samer Hijazi}, \bibinfo{person}{Rishi
  Kumar}, {and} \bibinfo{person}{Chris Rowen}.}
  \bibinfo{year}{2015}\natexlab{}.
\newblock \bibinfo{booktitle}{\emph{Using convolutional neural networks for
  image recognition}}.
\newblock \bibinfo{type}{{T}echnical {R}eport}. \bibinfo{institution}{Tech.
  Rep., 2015.[Online]. Available: http://ip. cadence.
  com/uploads/901/cnn-wp-pdf}.
\newblock


\bibitem[\protect\citeauthoryear{Hochreiter, Bengio, Frasconi, Schmidhuber,
  et~al\mbox{.}}{Hochreiter et~al\mbox{.}}{2001}]%
        {hochreiter2001gradient}
\bibfield{author}{\bibinfo{person}{Sepp Hochreiter}, \bibinfo{person}{Yoshua
  Bengio}, \bibinfo{person}{Paolo Frasconi}, \bibinfo{person}{J{\"u}rgen
  Schmidhuber}, {et~al\mbox{.}}} \bibinfo{year}{2001}\natexlab{}.
\newblock \bibinfo{title}{Gradient flow in recurrent nets: the difficulty of
  learning long-term dependencies}.
\newblock   (\bibinfo{year}{2001}).
\newblock


\bibitem[\protect\citeauthoryear{Hochreiter and Schmidhuber}{Hochreiter and
  Schmidhuber}{1997}]%
        {hochreiter1997long}
\bibfield{author}{\bibinfo{person}{Sepp Hochreiter} {and}
  \bibinfo{person}{J{\"u}rgen Schmidhuber}.} \bibinfo{year}{1997}\natexlab{}.
\newblock \showarticletitle{Long short-term memory}.
\newblock \bibinfo{journal}{\emph{Neural computation}} \bibinfo{volume}{9},
  \bibinfo{number}{8} (\bibinfo{year}{1997}), \bibinfo{pages}{1735--1780}.
\newblock


\bibitem[\protect\citeauthoryear{Huang, Kwiatkowska, Wang, and Wu}{Huang
  et~al\mbox{.}}{2017}]%
        {huang2017safety}
\bibfield{author}{\bibinfo{person}{Xiaowei Huang}, \bibinfo{person}{Marta
  Kwiatkowska}, \bibinfo{person}{Sen Wang}, {and} \bibinfo{person}{Min Wu}.}
  \bibinfo{year}{2017}\natexlab{}.
\newblock \showarticletitle{Safety verification of deep neural networks}. In
  \bibinfo{booktitle}{\emph{International Conference on Computer Aided
  Verification}}. Springer, \bibinfo{pages}{3--29}.
\newblock


\bibitem[\protect\citeauthoryear{Jain and Medsker}{Jain and Medsker}{1999}]%
        {rnn}
\bibfield{author}{\bibinfo{person}{L.~C. Jain} {and} \bibinfo{person}{L.~R.
  Medsker}.} \bibinfo{year}{1999}\natexlab{}.
\newblock \bibinfo{booktitle}{\emph{Recurrent Neural Networks: Design and
  Applications} (\bibinfo{edition}{1st} ed.)}.
\newblock \bibinfo{publisher}{CRC Press, Inc.}, \bibinfo{address}{Boca Raton,
  FL, USA}.
\newblock


\bibitem[\protect\citeauthoryear{Karpathy}{Karpathy}{[n. d.]}]%
        {cnn}
\bibfield{author}{\bibinfo{person}{Andrej Karpathy}.} \bibinfo{year}{[n.
  d.]}\natexlab{}.
\newblock \bibinfo{title}{Convolutional neural networks}.
\newblock
  \bibinfo{howpublished}{\url{http://cs231n.github.io/convolutional-networks/}}.
    (\bibinfo{year}{[n. d.]}).
\newblock


\bibitem[\protect\citeauthoryear{Katz, Barrett, Dill, Julian, and
  Kochenderfer}{Katz et~al\mbox{.}}{2017}]%
        {katz2017reluplex}
\bibfield{author}{\bibinfo{person}{Guy Katz}, \bibinfo{person}{Clark Barrett},
  \bibinfo{person}{David~L. Dill}, \bibinfo{person}{Kyle Julian}, {and}
  \bibinfo{person}{Mykel~J. Kochenderfer}.} \bibinfo{year}{2017}\natexlab{}.
\newblock \bibinfo{booktitle}{\emph{Reluplex: An Efficient SMT Solver for
  Verifying Deep Neural Networks}}.
\newblock \bibinfo{publisher}{Springer International Publishing},
  \bibinfo{address}{Cham}, \bibinfo{pages}{97--117}.
\newblock


\bibitem[\protect\citeauthoryear{Kos, Fischer, and Song}{Kos
  et~al\mbox{.}}{2017}]%
        {kos2017adversarial}
\bibfield{author}{\bibinfo{person}{Jernej Kos}, \bibinfo{person}{Ian Fischer},
  {and} \bibinfo{person}{Dawn Song}.} \bibinfo{year}{2017}\natexlab{}.
\newblock \showarticletitle{Adversarial examples for generative models}.
\newblock \bibinfo{journal}{\emph{arXiv preprint arXiv:1702.06832}}
  (\bibinfo{year}{2017}).
\newblock


\bibitem[\protect\citeauthoryear{Krizhevsky, Sutskever, and Hinton}{Krizhevsky
  et~al\mbox{.}}{2012}]%
        {krizhevsky2012imagenet}
\bibfield{author}{\bibinfo{person}{Alex Krizhevsky}, \bibinfo{person}{Ilya
  Sutskever}, {and} \bibinfo{person}{Geoffrey~E Hinton}.}
  \bibinfo{year}{2012}\natexlab{}.
\newblock \showarticletitle{Imagenet classification with deep convolutional
  neural networks}. In \bibinfo{booktitle}{\emph{Advances in neural information
  processing systems}}. \bibinfo{pages}{1097--1105}.
\newblock


\bibitem[\protect\citeauthoryear{Laskov et~al\mbox{.}}{Laskov
  et~al\mbox{.}}{2014}]%
        {laskov2014practical}
\bibfield{author}{\bibinfo{person}{Pavel Laskov} {et~al\mbox{.}}}
  \bibinfo{year}{2014}\natexlab{}.
\newblock \showarticletitle{Practical evasion of a learning-based classifier: A
  case study}. In \bibinfo{booktitle}{\emph{Security and Privacy (SP), 2014
  IEEE Symposium on}}. IEEE, \bibinfo{pages}{197--211}.
\newblock


\bibitem[\protect\citeauthoryear{Liu, Chen, Liu, and Song}{Liu
  et~al\mbox{.}}{2017}]%
        {liu2016delving}
\bibfield{author}{\bibinfo{person}{Yanpei Liu}, \bibinfo{person}{Xinyun Chen},
  \bibinfo{person}{Chang Liu}, {and} \bibinfo{person}{Dawn~Xiaodong Song}.}
  \bibinfo{year}{2017}\natexlab{}.
\newblock \showarticletitle{Delving into Transferable Adversarial Examples and
  Black-box Attacks}. In \bibinfo{booktitle}{\emph{International Conference on
  Learning Representations (ICLR)}}.
\newblock


\bibitem[\protect\citeauthoryear{McMinn}{McMinn}{2004}]%
        {mcminn2004search}
\bibfield{author}{\bibinfo{person}{Phil McMinn}.}
  \bibinfo{year}{2004}\natexlab{}.
\newblock \showarticletitle{Search-based software test data generation: a
  survey}.
\newblock \bibinfo{journal}{\emph{Software testing, Verification and
  reliability}} \bibinfo{volume}{14}, \bibinfo{number}{2}
  (\bibinfo{year}{2004}), \bibinfo{pages}{105--156}.
\newblock


\bibitem[\protect\citeauthoryear{Metzen, Genewein, Fischer, and
  Bischoff}{Metzen et~al\mbox{.}}{2017}]%
        {metzen2017detecting}
\bibfield{author}{\bibinfo{person}{Jan~Hendrik Metzen}, \bibinfo{person}{Tim
  Genewein}, \bibinfo{person}{Volker Fischer}, {and} \bibinfo{person}{Bastian
  Bischoff}.} \bibinfo{year}{2017}\natexlab{}.
\newblock \showarticletitle{On detecting adversarial perturbations}. In
  \bibinfo{booktitle}{\emph{International Conference on Learning
  Representations (ICLR)}}.
\newblock


\bibitem[\protect\citeauthoryear{Mitchell}{Mitchell}{1997}]%
        {Mitchell:1997:ML:541177}
\bibfield{author}{\bibinfo{person}{Thomas~M. Mitchell}.}
  \bibinfo{year}{1997}\natexlab{}.
\newblock \bibinfo{booktitle}{\emph{Machine Learning} (\bibinfo{edition}{1}
  ed.)}.
\newblock \bibinfo{publisher}{McGraw-Hill, Inc.}, \bibinfo{address}{New York,
  NY, USA}.
\newblock


\bibitem[\protect\citeauthoryear{Miyato, Dai, and Goodfellow}{Miyato
  et~al\mbox{.}}{2016}]%
        {miyato2016adversarial}
\bibfield{author}{\bibinfo{person}{Takeru Miyato}, \bibinfo{person}{Andrew~M
  Dai}, {and} \bibinfo{person}{Ian Goodfellow}.}
  \bibinfo{year}{2016}\natexlab{}.
\newblock \showarticletitle{Adversarial Training Methods for Semi-Supervised
  Text Classification}. In \bibinfo{booktitle}{\emph{Proceedings of the
  International Conference on Learning Representations (ICLR)}}.
\newblock


\bibitem[\protect\citeauthoryear{Murphy, Kaiser, Hu, and Wu}{Murphy
  et~al\mbox{.}}{2008}]%
        {murphy2008properties}
\bibfield{author}{\bibinfo{person}{Christian Murphy}, \bibinfo{person}{Gail~E
  Kaiser}, \bibinfo{person}{Lifeng Hu}, {and} \bibinfo{person}{Leon Wu}.}
  \bibinfo{year}{2008}\natexlab{}.
\newblock \showarticletitle{Properties of Machine Learning Applications for Use
  in Metamorphic Testing.}. In \bibinfo{booktitle}{\emph{SEKE}},
  Vol.~\bibinfo{volume}{8}. \bibinfo{pages}{867--872}.
\newblock


\bibitem[\protect\citeauthoryear{Nair and Hinton}{Nair and Hinton}{2010}]%
        {nair2010rectified}
\bibfield{author}{\bibinfo{person}{Vinod Nair} {and}
  \bibinfo{person}{Geoffrey~E Hinton}.} \bibinfo{year}{2010}\natexlab{}.
\newblock \showarticletitle{Rectified linear units improve restricted boltzmann
  machines}. In \bibinfo{booktitle}{\emph{Proceedings of the 27th international
  conference on machine learning (ICML-10)}}. \bibinfo{pages}{807--814}.
\newblock


\bibitem[\protect\citeauthoryear{Narodytska and Kasiviswanathan}{Narodytska and
  Kasiviswanathan}{2016}]%
        {narodytska2016simple}
\bibfield{author}{\bibinfo{person}{Nina Narodytska} {and}
  \bibinfo{person}{Shiva~Prasad Kasiviswanathan}.}
  \bibinfo{year}{2016}\natexlab{}.
\newblock \showarticletitle{Simple black-box adversarial perturbations for deep
  networks}. In \bibinfo{booktitle}{\emph{Workshop on Adversarial Training,
  NIPS 2016}}.
\newblock


\bibitem[\protect\citeauthoryear{Nguyen, Yosinski, and Clune}{Nguyen
  et~al\mbox{.}}{2015}]%
        {nguyen2015deep}
\bibfield{author}{\bibinfo{person}{Anh Nguyen}, \bibinfo{person}{Jason
  Yosinski}, {and} \bibinfo{person}{Jeff Clune}.}
  \bibinfo{year}{2015}\natexlab{}.
\newblock \showarticletitle{Deep neural networks are easily fooled: High
  confidence predictions for unrecognizable images}. In
  \bibinfo{booktitle}{\emph{Proceedings of the IEEE Conference on Computer
  Vision and Pattern Recognition}}. \bibinfo{pages}{427--436}.
\newblock


\bibitem[\protect\citeauthoryear{Papernot and McDaniel}{Papernot and
  McDaniel}{2017}]%
        {papernot2017extending}
\bibfield{author}{\bibinfo{person}{Nicolas Papernot} {and}
  \bibinfo{person}{Patrick McDaniel}.} \bibinfo{year}{2017}\natexlab{}.
\newblock \showarticletitle{Extending Defensive Distillation}.
\newblock \bibinfo{journal}{\emph{arXiv preprint arXiv:1705.05264}}
  (\bibinfo{year}{2017}).
\newblock


\bibitem[\protect\citeauthoryear{Papernot, McDaniel, Goodfellow, Jha, Celik,
  and Swami}{Papernot et~al\mbox{.}}{2017}]%
        {papernot2017practical}
\bibfield{author}{\bibinfo{person}{Nicolas Papernot}, \bibinfo{person}{Patrick
  McDaniel}, \bibinfo{person}{Ian Goodfellow}, \bibinfo{person}{Somesh Jha},
  \bibinfo{person}{Z~Berkay Celik}, {and} \bibinfo{person}{Ananthram Swami}.}
  \bibinfo{year}{2017}\natexlab{}.
\newblock \showarticletitle{Practical black-box attacks against machine
  learning}. In \bibinfo{booktitle}{\emph{Proceedings of the 2017 ACM on Asia
  Conference on Computer and Communications Security}}. ACM,
  \bibinfo{pages}{506--519}.
\newblock


\bibitem[\protect\citeauthoryear{Papernot, McDaniel, Jha, Fredrikson, Celik,
  and Swami}{Papernot et~al\mbox{.}}{2016a}]%
        {papernot2016limitations}
\bibfield{author}{\bibinfo{person}{Nicolas Papernot}, \bibinfo{person}{Patrick
  McDaniel}, \bibinfo{person}{Somesh Jha}, \bibinfo{person}{Matt Fredrikson},
  \bibinfo{person}{Z~Berkay Celik}, {and} \bibinfo{person}{Ananthram Swami}.}
  \bibinfo{year}{2016}\natexlab{a}.
\newblock \showarticletitle{The limitations of deep learning in adversarial
  settings}. In \bibinfo{booktitle}{\emph{2016 IEEE European Symposium on
  Security and Privacy (EuroS\&P)}}. IEEE, \bibinfo{pages}{372--387}.
\newblock


\bibitem[\protect\citeauthoryear{Papernot, McDaniel, Swami, and
  Harang}{Papernot et~al\mbox{.}}{2016b}]%
        {papernot2016crafting}
\bibfield{author}{\bibinfo{person}{Nicolas Papernot}, \bibinfo{person}{Patrick
  McDaniel}, \bibinfo{person}{Ananthram Swami}, {and} \bibinfo{person}{Richard
  Harang}.} \bibinfo{year}{2016}\natexlab{b}.
\newblock \showarticletitle{Crafting adversarial input sequences for recurrent
  neural networks}. In \bibinfo{booktitle}{\emph{Military Communications
  Conference, MILCOM 2016-2016 IEEE}}. IEEE, \bibinfo{pages}{49--54}.
\newblock


\bibitem[\protect\citeauthoryear{Papernot, McDaniel, Wu, Jha, and
  Swami}{Papernot et~al\mbox{.}}{2016c}]%
        {papernot2016distillation}
\bibfield{author}{\bibinfo{person}{Nicolas Papernot}, \bibinfo{person}{Patrick
  McDaniel}, \bibinfo{person}{Xi Wu}, \bibinfo{person}{Somesh Jha}, {and}
  \bibinfo{person}{Ananthram Swami}.} \bibinfo{year}{2016}\natexlab{c}.
\newblock \showarticletitle{Distillation as a defense to adversarial
  perturbations against deep neural networks}. In
  \bibinfo{booktitle}{\emph{Security and Privacy (SP), 2016 IEEE Symposium
  on}}. IEEE, \bibinfo{pages}{582--597}.
\newblock


\bibitem[\protect\citeauthoryear{P{\u{a}}s{\u{a}}reanu and
  Visser}{P{\u{a}}s{\u{a}}reanu and Visser}{2009}]%
        {puasuareanu2009survey}
\bibfield{author}{\bibinfo{person}{Corina~S P{\u{a}}s{\u{a}}reanu} {and}
  \bibinfo{person}{Willem Visser}.} \bibinfo{year}{2009}\natexlab{}.
\newblock \showarticletitle{A survey of new trends in symbolic execution for
  software testing and analysis}.
\newblock \bibinfo{journal}{\emph{International Journal on Software Tools for
  Technology Transfer (STTT)}} \bibinfo{volume}{11}, \bibinfo{number}{4}
  (\bibinfo{year}{2009}), \bibinfo{pages}{339--353}.
\newblock


\bibitem[\protect\citeauthoryear{Pei, Cao, Yang, and Jana}{Pei
  et~al\mbox{.}}{2017}]%
        {pei2017deepxplore}
\bibfield{author}{\bibinfo{person}{Kexin Pei}, \bibinfo{person}{Yinzhi Cao},
  \bibinfo{person}{Junfeng Yang}, {and} \bibinfo{person}{Suman Jana}.}
  \bibinfo{year}{2017}\natexlab{}.
\newblock \showarticletitle{DeepXplore: Automated Whitebox Testing of Deep
  Learning Systems}.
\newblock \bibinfo{journal}{\emph{arXiv preprint arXiv:1705.06640}}
  (\bibinfo{year}{2017}).
\newblock


\bibitem[\protect\citeauthoryear{Pulina and Tacchella}{Pulina and
  Tacchella}{2010}]%
        {pulina2010abstraction}
\bibfield{author}{\bibinfo{person}{Luca Pulina} {and} \bibinfo{person}{Armando
  Tacchella}.} \bibinfo{year}{2010}\natexlab{}.
\newblock \showarticletitle{An abstraction-refinement approach to verification
  of artificial neural networks}. In \bibinfo{booktitle}{\emph{Computer Aided
  Verification}}. Springer, \bibinfo{pages}{243--257}.
\newblock


\bibitem[\protect\citeauthoryear{Rumelhart, Hinton, and Williams}{Rumelhart
  et~al\mbox{.}}{1988}]%
        {rumelhart1988learning}
\bibfield{author}{\bibinfo{person}{David~E Rumelhart},
  \bibinfo{person}{Geoffrey~E Hinton}, {and} \bibinfo{person}{Ronald~J
  Williams}.} \bibinfo{year}{1988}\natexlab{}.
\newblock \showarticletitle{Learning representations by back-propagating
  errors}.
\newblock \bibinfo{journal}{\emph{Cognitive modeling}} \bibinfo{volume}{5},
  \bibinfo{number}{3} (\bibinfo{year}{1988}), \bibinfo{pages}{1}.
\newblock


\bibitem[\protect\citeauthoryear{Sculley, Holt, Golovin, Davydov, Phillips,
  Ebner, Chaudhary, and Young}{Sculley et~al\mbox{.}}{2014}]%
        {sculley2014machine}
\bibfield{author}{\bibinfo{person}{D. Sculley}, \bibinfo{person}{Gary Holt},
  \bibinfo{person}{Daniel Golovin}, \bibinfo{person}{Eugene Davydov},
  \bibinfo{person}{Todd Phillips}, \bibinfo{person}{Dietmar Ebner},
  \bibinfo{person}{Vinay Chaudhary}, {and} \bibinfo{person}{Michael Young}.}
  \bibinfo{year}{2014}\natexlab{}.
\newblock \showarticletitle{Machine Learning: The High Interest Credit Card of
  Technical Debt}.
\newblock


\bibitem[\protect\citeauthoryear{Shaham, Yamada, and Negahban}{Shaham
  et~al\mbox{.}}{2015}]%
        {shaham2015understanding}
\bibfield{author}{\bibinfo{person}{Uri Shaham}, \bibinfo{person}{Yutaro
  Yamada}, {and} \bibinfo{person}{Sahand Negahban}.}
  \bibinfo{year}{2015}\natexlab{}.
\newblock \showarticletitle{Understanding adversarial training: Increasing
  local stability of neural nets through robust optimization}.
\newblock \bibinfo{journal}{\emph{arXiv preprint arXiv:1511.05432}}
  (\bibinfo{year}{2015}).
\newblock


\bibitem[\protect\citeauthoryear{Sharif, Bhagavatula, Bauer, and Reiter}{Sharif
  et~al\mbox{.}}{2016}]%
        {sharif2016accessorize}
\bibfield{author}{\bibinfo{person}{Mahmood Sharif}, \bibinfo{person}{Sruti
  Bhagavatula}, \bibinfo{person}{Lujo Bauer}, {and} \bibinfo{person}{Michael~K
  Reiter}.} \bibinfo{year}{2016}\natexlab{}.
\newblock \showarticletitle{Accessorize to a crime: Real and stealthy attacks
  on state-of-the-art face recognition}. In
  \bibinfo{booktitle}{\emph{Proceedings of the 2016 ACM SIGSAC Conference on
  Computer and Communications Security}}. ACM, \bibinfo{pages}{1528--1540}.
\newblock


\bibitem[\protect\citeauthoryear{Spearman}{Spearman}{1904}]%
        {spearman1904proof}
\bibfield{author}{\bibinfo{person}{Charles Spearman}.}
  \bibinfo{year}{1904}\natexlab{}.
\newblock \showarticletitle{The proof and measurement of association between
  two things}.
\newblock \bibinfo{journal}{\emph{The American journal of psychology}}
  \bibinfo{volume}{15}, \bibinfo{number}{1} (\bibinfo{year}{1904}),
  \bibinfo{pages}{72--101}.
\newblock


\bibitem[\protect\citeauthoryear{Steinhardt, Koh, and Liang}{Steinhardt
  et~al\mbox{.}}{2017}]%
        {steinhardt2017certified}
\bibfield{author}{\bibinfo{person}{Jacob Steinhardt}, \bibinfo{person}{Pang~Wei
  Koh}, {and} \bibinfo{person}{Percy Liang}.} \bibinfo{year}{2017}\natexlab{}.
\newblock \showarticletitle{Certified Defenses for Data Poisoning Attacks}.
\newblock \bibinfo{journal}{\emph{arXiv preprint arXiv:1706.03691}}
  (\bibinfo{year}{2017}).
\newblock


\bibitem[\protect\citeauthoryear{Szegedy, Zaremba, Sutskever, Bruna, Erhan,
  Goodfellow, and Fergus}{Szegedy et~al\mbox{.}}{2014}]%
        {szegedy2013intriguing}
\bibfield{author}{\bibinfo{person}{C. Szegedy}, \bibinfo{person}{W. Zaremba},
  \bibinfo{person}{I. Sutskever}, \bibinfo{person}{J. Bruna},
  \bibinfo{person}{D. Erhan}, \bibinfo{person}{I. Goodfellow}, {and}
  \bibinfo{person}{R. Fergus}.} \bibinfo{year}{2014}\natexlab{}.
\newblock \showarticletitle{Intriguing properties of neural networks}. In
  \bibinfo{booktitle}{\emph{International Conference on Learning
  Representations (ICLR)}}.
\newblock


\bibitem[\protect\citeauthoryear{{Theano Development Team}}{{Theano Development
  Team}}{2016}]%
        {2016arXiv160502688short}
\bibfield{author}{\bibinfo{person}{{Theano Development Team}}.}
  \bibinfo{year}{2016}\natexlab{}.
\newblock \showarticletitle{{Theano: A {Python} framework for fast computation
  of mathematical expressions}}.
\newblock \bibinfo{journal}{\emph{arXiv e-prints}}
  \bibinfo{volume}{abs/1605.02688} (\bibinfo{date}{May} \bibinfo{year}{2016}).
\newblock
\urldef\tempurl%
\url{http://arxiv.org/abs/1605.02688}
\showURL{%
\tempurl}


\bibitem[\protect\citeauthoryear{Wang, Wang, Zheng, and Zhao}{Wang
  et~al\mbox{.}}{2014}]%
        {wang2014man}
\bibfield{author}{\bibinfo{person}{Gang Wang}, \bibinfo{person}{Tianyi Wang},
  \bibinfo{person}{Haitao Zheng}, {and} \bibinfo{person}{Ben~Y Zhao}.}
  \bibinfo{year}{2014}\natexlab{}.
\newblock \showarticletitle{Man vs. Machine: Practical Adversarial Detection of
  Malicious Crowdsourcing Workers.}. In \bibinfo{booktitle}{\emph{USENIX
  Security Symposium}}. \bibinfo{pages}{239--254}.
\newblock


\bibitem[\protect\citeauthoryear{Wilber, Shmatikov, and Belongie}{Wilber
  et~al\mbox{.}}{2016}]%
        {wilber2016can}
\bibfield{author}{\bibinfo{person}{Michael~J Wilber}, \bibinfo{person}{Vitaly
  Shmatikov}, {and} \bibinfo{person}{Serge Belongie}.}
  \bibinfo{year}{2016}\natexlab{}.
\newblock \showarticletitle{Can we still avoid automatic face detection?}. In
  \bibinfo{booktitle}{\emph{Applications of Computer Vision (WACV), 2016 IEEE
  Winter Conference on}}. IEEE, \bibinfo{pages}{1--9}.
\newblock


\bibitem[\protect\citeauthoryear{Witten, Frank, Hall, and Pal}{Witten
  et~al\mbox{.}}{2016}]%
        {witten2016data}
\bibfield{author}{\bibinfo{person}{Ian~H Witten}, \bibinfo{person}{Eibe Frank},
  \bibinfo{person}{Mark~A Hall}, {and} \bibinfo{person}{Christopher~J Pal}.}
  \bibinfo{year}{2016}\natexlab{}.
\newblock \bibinfo{booktitle}{\emph{Data Mining: Practical machine learning
  tools and techniques}}.
\newblock \bibinfo{publisher}{Morgan Kaufmann}.
\newblock


\bibitem[\protect\citeauthoryear{Xie, Ho, Murphy, Kaiser, Xu, and Chen}{Xie
  et~al\mbox{.}}{2009}]%
        {xie2009application}
\bibfield{author}{\bibinfo{person}{Xiaoyuan Xie}, \bibinfo{person}{Joshua Ho},
  \bibinfo{person}{Christian Murphy}, \bibinfo{person}{Gail Kaiser},
  \bibinfo{person}{Baowen Xu}, {and} \bibinfo{person}{Tsong~Yueh Chen}.}
  \bibinfo{year}{2009}\natexlab{}.
\newblock \showarticletitle{Application of metamorphic testing to supervised
  classifiers}. In \bibinfo{booktitle}{\emph{Quality Software, 2009. QSIC'09.
  9th International Conference on}}. IEEE, \bibinfo{pages}{135--144}.
\newblock


\bibitem[\protect\citeauthoryear{Xu, Evans, and Qi}{Xu et~al\mbox{.}}{2017}]%
        {xu2017feature}
\bibfield{author}{\bibinfo{person}{Weilin Xu}, \bibinfo{person}{David Evans},
  {and} \bibinfo{person}{Yanjun Qi}.} \bibinfo{year}{2017}\natexlab{}.
\newblock \showarticletitle{Feature Squeezing: Detecting Adversarial Examples
  in Deep Neural Networks}.
\newblock \bibinfo{journal}{\emph{arXiv preprint arXiv:1704.01155}}
  (\bibinfo{year}{2017}).
\newblock


\bibitem[\protect\citeauthoryear{Xu, Qi, and Evans}{Xu et~al\mbox{.}}{2016}]%
        {xu2016automatically}
\bibfield{author}{\bibinfo{person}{Weilin Xu}, \bibinfo{person}{Yanjun Qi},
  {and} \bibinfo{person}{David Evans}.} \bibinfo{year}{2016}\natexlab{}.
\newblock \showarticletitle{Automatically evading classifiers}. In
  \bibinfo{booktitle}{\emph{Proceedings of the 2016 Network and Distributed
  Systems Symposium}}.
\newblock


\bibitem[\protect\citeauthoryear{Zheng, Song, Leung, and Goodfellow}{Zheng
  et~al\mbox{.}}{2016}]%
        {zheng2016improving}
\bibfield{author}{\bibinfo{person}{Stephan Zheng}, \bibinfo{person}{Yang Song},
  \bibinfo{person}{Thomas Leung}, {and} \bibinfo{person}{Ian Goodfellow}.}
  \bibinfo{year}{2016}\natexlab{}.
\newblock \showarticletitle{Improving the robustness of deep neural networks
  via stability training}. In \bibinfo{booktitle}{\emph{Proceedings of the IEEE
  Conference on Computer Vision and Pattern Recognition}}.
  \bibinfo{pages}{4480--4488}.
\newblock


\bibitem[\protect\citeauthoryear{Zhou, Huang, Tse, Yang, Huang, and Chen}{Zhou
  et~al\mbox{.}}{2004}]%
        {zhou2004metamorphic}
\bibfield{author}{\bibinfo{person}{Zhi~Quan Zhou}, \bibinfo{person}{DH Huang},
  \bibinfo{person}{TH Tse}, \bibinfo{person}{Zongyuan Yang},
  \bibinfo{person}{Haitao Huang}, {and} \bibinfo{person}{TY Chen}.}
  \bibinfo{year}{2004}\natexlab{}.
\newblock \showarticletitle{Metamorphic testing and its applications}. In
  \bibinfo{booktitle}{\emph{Proceedings of the 8th International Symposium on
  Future Software Technology (ISFST 2004)}}. \bibinfo{pages}{346--351}.
\newblock


\end{thebibliography}



\end{document}